\documentclass[12pt,preprint]{aastex}

\shorttitle{A grid of 200,000 YSO SEDs}
\shortauthors{Robitaille et al.}

\newcommand{\agestar}{t_{\star}}

\newcommand{\mstar}{M_{\star}}
\newcommand{\rstar}{R_{\star}}
\newcommand{\tstar}{T_{\star}}

\newcommand{\mdote}{\dot{M}_{\rm env}}
\newcommand{\rmaxe}{R_{\rm env}^{\rm max}}
\newcommand{\rmine}{R_{\rm env}^{\rm min}}

\newcommand{\mdisk}{M_{\rm disk}}
\newcommand{\mdotdisk}{\dot{M}_{\rm disk}}

\newcommand{\rmaxd}{R_{\rm disk}^{\rm max}}
\newcommand{\rmind}{R_{\rm disk}^{\rm min}}

\newcommand{\rsub}{R_{\rm sub}}

\newcommand{\rhoconst}{\rho_{\rm cavity}}
\newcommand{\thetacav}{\theta_{\rm cavity}}

\newcommand{\rhoamb}{\rho_{\rm ambient}}

\newcommand{\rc}{R_{\rm c}}

\newcommand{\zmin}{z_{\rm factor}}

\newcommand{\alpharef}{$\alpha_{[K\,\&\,\rm{MIPS}\,24]}$\,}
\newcommand{\alphaend}{$\alpha_{[\lambda_1\,\&\,\lambda_2]}$\,}
\newcommand{\alphaall}{$\alpha_{[\lambda_1\,\rightarrow\,\lambda_2]}$\,}

\newcommand{\alpharefns}{$\alpha_{[K\,\&\,\rm{MIPS}\,24]}$}
\newcommand{\alphaendns}{$\alpha_{[\lambda_1\,\&\,\lambda_2]}$}
\newcommand{\alphaallns}{$\alpha_{[\lambda_1\,\rightarrow\,\lambda_2]}$}

\newcommand{\msun}{M_{\odot}}

\newcommand{\mips}{MIPS 24\,$\mu$m~}
\newcommand{\mipsns}{MIPS 24\,$\mu$m}

\newcommand{\microns}{\,$\mu$m~}
\newcommand{\micronsns}{\,$\mu$m}

\newcommand{\hh}{{\rm H_2}}

\newcommand{\av}{A_{\rm V}}
\newcommand{\rv}{R_{\rm V}}

\begin{document}


\title{Interpreting Spectral Energy Distributions from Young Stellar Objects. I. A grid of 200,000 YSO model SEDs.}

\author{Thomas P. Robitaille\altaffilmark{1}}
\author{Barbara A. Whitney\altaffilmark{2}}
\author{Remy Indebetouw\altaffilmark{3}}
\author{Kenneth Wood\altaffilmark{1}}
\author{Pia Denzmore\altaffilmark{4}}

\altaffiltext{1}{SUPA, School of Physics and Astronomy, University of St Andrews, North 
Haugh, St Andrews, KY16 9SS, United Kingdom; tr9@st-andrews.ac.uk, kw25@st-andrews.ac.uk}
\altaffiltext{2}{Space Science Institute, 4750 Walnut St. Suite 205, Boulder, CO 80301, USA; bwhitney@spacescience.org}
\altaffiltext{3}{University of Virginia, Astronomy Dept., P.O. Box 3818, Charlottesville, VA, 22903-0818; remy@virginia.edu}
\altaffiltext{4}{Physics and Astronomy Department, Rice University, Houston, TX, USA; piadenz@rice.edu}


\begin{abstract}
We present a grid of radiation transfer models of axisymmetric young stellar objects (YSOs), covering a wide range of stellar masses (from $0.1$\,$\msun$ to $50$\,$\msun$) and evolutionary stages (from the early envelope infall stage to the late disk-only stage).
The grid consists of 20,000 YSO models, with spectral energy distributions (SEDs) and polarization spectra computed at ten viewing angles for each model, resulting in a total of 200,000 SEDs.
We made a careful assessment of the theoretical and observational constraints on the physical conditions of disks and envelopes in YSOs, and have attempted to fully span the corresponding regions in parameter space. 
These models are publicly available on a dedicated WWW server\,\footnote{http://www.astro.wisc.edu/protostars}.
In this paper we summarize the main features of our models, as well as the range of parameters explored.
Having a large grid covering reasonable regions of parameter space allows us to shed light on many trends in near- and mid-IR observations of YSOs (such as changes in the spectral indices and colors of their SEDs), linking them with physical parameters (such as disk and infalling envelope parameters).
In particular we examine the dependence of the spectral indices of the model SEDs on envelope accretion rate and disk mass.
In addition, we show variations of spectral indices with stellar temperature, disk inner radius, and disk flaring power for a subset of disk-only models.
We also examine how changing the wavelength range of data used to calculate spectral indices affects their values.
We show sample color-color plots of the entire grid as well as simulated clusters at various distances with typical {\it Spitzer Space Telescope} sensitivities.
We find that young embedded sources generally occupy a large region of color-color space due to inclination and stellar temperature effects.
Disk sources occupy a smaller region of color-color space, but overlap substantially with the region occupied by embedded sources, especially in the near- and mid-IR.
We identify regions in color-color space where our models indicate that only sources at a given evolutionary stage should lie.
We find that while near-IR (such as JHK) and mid-IR (such as IRAC) fluxes are useful in discriminating between stars and YSOs, and are useful for identifying very young sources, the addition of longer wavelength data such as \mips is extremely valuable for determining the evolutionary stage of YSOs.
\end{abstract}

\keywords{astronomical data bases: miscellaneous --- circumstellar matter --- infrared: stars --- radiative transfer --- stars: formation --- stars: pre-main-sequence --- polarization}


\section{Introduction}
\label{s:intro}
An explosion of infrared data on star formation regions is being collected by the {\it Spitzer Space Telescope} and will continue to be collected by future missions such as the {\it Herschel Space Observatory} and the {\it James Webb Space Telescope}.
Despite its modest size ($0.85$\,m), the {\it Spitzer Space Telescope} is much more sensitive and has higher spatial resolution than previous infrared observatories \citep{werner04}.
It is also very efficient, and has mapped a substantial fraction of the star formation regions in the Galaxy using the IRAC \citep[3.6, 4.5, 5.8, and 8\micronsns;][]{fazio04} and MIPS \citep[24, 70, and 160\micronsns;][]{rieke04} instruments.
For example, the GLIMPSE \citep{benjamin03} and MIPSGAL (PI Carey) Legacy surveys are mapping
over 150 square degrees of the inner Galactic plane;
the Cores-to-Disks (c2d) Legacy project has surveyed five nearby large molecular clouds \citep{evans03}; and several projects are surveying the Orion molecular cloud \citep{megeath_iau}, the Taurus molecular cloud \citep{hartmann05}, the LMC \citep{chu_letter,meixner06} and SMC, and many other well-known star formation regions \citep[e.g][]{gutermuth04,megeath_etacam,allen_ppv,sicilia06}.
These data can be combined with other surveys, such as the 2MASS near-infrared all-sky survey
\citep{skrutskie06}, expanding the wavelength coverage of the observed YSOs.

Our particular goal is to characterize YSOs in the Galaxy using the GLIMPSE and MIPSGAL surveys, as well as in the LMC using the SAGE survey, and to determine the timescales of various evolutionary stages as a function of stellar mass and location in the Galaxy.
Ultimately, we hope to provide an independent estimate of the star formation rate and efficiency in the Galaxy and the LMC.

To help analyze the SEDs of YSOs we developed radiation transfer models \citep{whitney03p2,whitney03p1,whitney04}. 
The well-known classification for low-mass YSOs uses the spectral index $\alpha$ (the slope of $\log_{10}{\lambda F_\lambda}$ vs $\log_{10}{\lambda}$ longward of $2$\micronsns) to classify a source as embedded ($\alpha > 0$; Class~I), a disk source ($-2 < \alpha < 0$; Class~II) or a source with an optically thin or no disk ($\alpha < -2$; Class~III) \citep{lada87}. 
However, our models indicate that the spectral index and the colors of a YSO in certain wavelength ranges are not always directly related to its evolutionary stage.
For example, an edge-on embedded protostar can have a decreasing slope in the narrow wavelength range of $2-10$\microns if the flux is dominated by scattered light 
(edge-on disks exhibit similar behavior, \citealt{wood02_hh30}, \citealt{grosso03}).
In addition, the temperature of the central source also affects the colors in this wavelength range, as does the location of the inner radius of the disk:  hot stellar sources and large inner disk holes can produce red colors in a star+disk source (see Section~\ref{s:color}).

In an attempt to improve our physical understanding gained from interpreting the SEDs of the many sources found in these surveys, many of which only
have JHK, IRAC, and \mips data, we plan to fit observed SEDs using a large grid of pre-computed model SEDs.
This grid attempts to encompass a large range of stellar masses and YSO evolutionary stages.
We sampled the model parameters based on both theory and observations (see Section \ref{s:params}).
This paper describes our publicly available grid of models.
A companion paper (Robitaille et al. 2006, in preparation) describes the method used to fit observed SEDs using the grid of models.
The advantages of fitting pre-computed SEDs to data, even in the fairly narrow wavelength range mentioned above, are that 1) one makes use of all available data simultaneously without loss of information 2) the uniqueness or non-uniqueness of a fit is immediately apparent from the range of model parameters that can fit a given SED, and 3) it is an efficient technique when used with large datasets as the model SEDs do not have to be computed for each source. 
In Section~\ref{s:models} we describe the grid of models.
Section~\ref{s:results} shows results from the grid, including sample SEDs, polarization spectra, and color-color plots, as well as an analysis of spectral index and color-color plot classifications.
Finally, in Section~\ref{s:conclusion} we make concluding remarks.

\section{The grid of models}
\label{s:models}
\subsection{The radiation transfer code}

\subsubsection{Brief description of the code}

The Monte Carlo radiation transfer code used for this grid of models includes non-isotropic scattering, polarization, and thermal emission from dust in a spherical-polar grid, solving for the temperature using the method of \citet{bjorkman01}.
This code is publicly available\,\footnote{http://www.astro.wisc.edu/protostars}.
The circumstellar geometry consists of a rotationally-flattened infalling envelope \citep{ulrich76,terebey84}, bipolar cavities \citep{whitney93,whitney03p2} and a flared accretion disk \citep{shakura73,pringle81,kenyon87,chiang97,dalessio98,dullemond01}.
The luminosity sources include the central star and disk accretion.
The code and the model geometries are described in detail in \citet{whitney03p2,whitney03p1}. 

As discussed in Section~\ref{s:params}, the various model parameters are sampled in order to produce a range of evolutionary stages.
For example, the envelope accretion rate decreases over time, the bipolar cavities become wider, the dust in the cavities less dense, and the disk radius increases during the early accretion from the envelope.

\subsubsection{Dust grain models}

Our grain models contain a mixture of astronomical silicates and graphite in solar abundance, using the optical constants of \citet{laor93}. 
The optical properties are averaged over the size distribution and composition. 
Thus we do not separate the heating and emission properties of different grain sizes or composition.
This could affect the thermal and chemical properties in the inner regions of the disk \citep{wolf03_mix} but its affect on the SED is expected to be relatively minor \citep{wolf03_mix,carciofi04}.

The grain properties vary with location in the disk and envelope as follows: 
the densest regions of the disk ($n_{\rm H_2} > 10^{10}$ cm$^{-3}$) use a grain model
with a size distribution that decays exponentially for sizes larger than 50\microns and extends up to $1$\,mm; this grain model fits the SED of the HH30 disk \citep{wood02_hh30}. 
This is the same as the ``Disk midplane'' grain model described in Table~3 of \citet{whitney03p2}.
The rest of the circumstellar geometry uses a grain size distribution with an average particle size slightly larger than the diffuse ISM, and a ratio of total-to-selective extinction $\rv=3.6$.
This is the ``KMH'' model \citep*{kmh94}, referred to as the ``Outflow'' model in Table~3 of \citet{whitney03p2}.
Dust grains in embedded regions of Taurus show evidence for further grain growth, with $\rv\sim4$ in the densest regions \citep{whittet01}, but recent models of near-IR images of Taurus protostars show that larger grain models are not well-distinguished from ISM grains \citep{wolf03,stark06,gramajo06}.
Grains in molecular clouds also show evidence for ice coatings \citep[e.g.][]{boogert04,knez05}, which are not included in our models.

\subsubsection{Stellar photospheres}

The spectrum of the central source for each model is dependent mainly on its temperature and to a lesser extent its surface gravity.
For stellar temperatures below $10,000$\,K, we used model stellar photospheres from \citet{brott05}, while for stellar temperatures above $10,000$\,K we used model stellar photospheres from \citet{kurucz93}.
In both cases we assumed solar metallicity.
For each YSO model, we interpolated the stellar photospheres to the relevant temperature and surface gravity.

\subsubsection{Model parameters}

Technically speaking the set of models we present in this paper does not form a `grid', since the parameters are randomly sampled within ranges. However, we will refer to the set of models as a `model grid', since this is a useful descriptive term.

The 14 model parameters are shown in Table~\ref{t:parameters}.  
Fortunately, only a few parameters are important at a given evolutionary stage.
For example in the youngest stages, the disk is hidden beneath the envelope: the disk inner radius, accretion rate, and to a lesser extent disk mass are the main disk parameters that affect the 1-8\microns fluxes.
The presence of an inner disk is required to produce mid-IR flux and to obscure the central source at edge-on viewing angles, but the dust properties in the disk and the amount of flaring do not have an important effect on the mid-IR SED.  
At these young stages, the most important parameters are the envelope accretion rate, the opening angle of the bipolar cavities, the inclination to the line of sight, the disk/envelope inner radius, the stellar temperature, and to a lesser extent the disk mass.
At later stages, when the envelope has mostly dispersed, the most relevant parameters for the SED are the disk inner radius, accretion rate, mass, and flaring (or dust settling) \citep{kenyon87,lada92,chiang97,dalessio98,furlan05}.

The details of the parameter sampling are given in Section~\ref{s:params}. 
It is important to note at this point that we are not suggesting our own model of evolution for YSOs.
Instead, our aim is to provide model SEDs for stages of evolution that have been suggested by theory or observations.
Furthermore, we do not follow several objects throughout their evolution, but instead we sample possible evolutionary stages and stellar masses randomly.

\subsubsection{Output of the radiation transfer code}
\label{s:output}
The output from the code presented here consists of flux and polarization spectra for 250 wavelengths (from $0.01$ to $5,000$\micronsns), computed at 10 viewing angles (from pole-on to edge-on in equal intervals of cosine of the inclination) and in 50 different circular apertures (with radii from $100$ to $100,000$\,AU).
Since each one of the 20,000 models produces an SED (flux vs. wavelength) for each viewing angle and aperture, our grid of models contains 10 million SEDs.
The code can produce images at specific viewing angles, but we did not compute these in the current grid as doing so would increase the CPU time required.
However, the 50 apertures from $100-100,000$\,AU amount to a spherically averaged intensity profile for each viewing angle of each model.

We have convolved all our models with a large number of common filter bandpasses ranging from optical to sub-mm wavelengths, including for example optical (e.g. UBVRI), near-IR (e.g. 2MASS JHK), mid- and far-IR (e.g. IRAC, MSX, IRAS, MIPS), and sub-mm (e.g. Scuba $450$ and $850$\micronsns) filters.
We will expand the range of filters used to convolve the SEDs as requested by users.
The polarization spectra have a lower signal-to-noise than the SEDs, so we smooth them before convolving with broadband filter profiles.

\begin{figure}
\epsscale{0.5}
\plotone{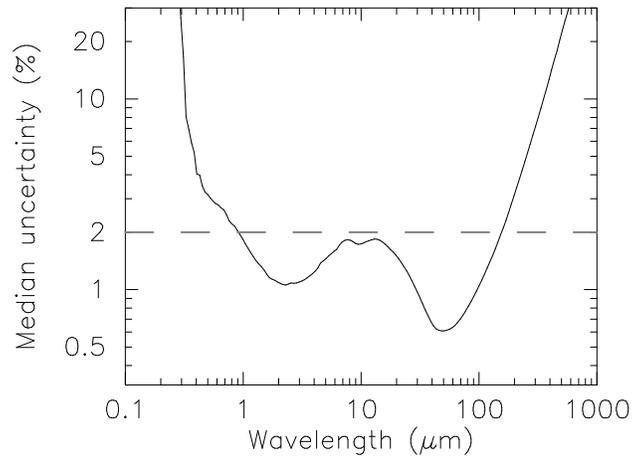}
\caption{The median fractional (one-sigma) error of the flux values for all 200,000 SEDs. For zero fluxes, a fractional error of 100\% was used. The dashed line indicates the level of 2\% uncertainty, and shows that the median error of our models is less than this in the range $1-100$\micronsns.\label{f:quality}}
\end{figure}

For the grid of models presented here, we ran each model with 20,000,000 `photons' \citep[energy packets,][]{bjorkman01}.
This produces SEDs with good signal-to-noise ratios for wavelengths spanning $1-100$\microns (however, we note that the signal-to noise may also be good outside this range).
The wavelength range at which a model SED has a good signal-to-noise depends on the evolutionary stage of the YSO: SEDs may be noisy at wavelengths shorter than $1$\microns, but with a good signal-to noise at sub-mm wavelengths for young embedded sources, or be noisy and at wavelengths longer than about $100$\microns and with a good signal-to-noise at optical and near-UV wavelengths for low mass disks.
Figure~\ref{f:quality} shows the median noise levels as a function of wavelength (using the SEDs measured in the largest aperture). 
We provide estimated uncertainties on our SEDs so that they are still usable in most wavelength ranges.
The SEDs can be re-binned to a lower wavelength resolution to reduce these uncertainties. 
Future versions of the grid will include higher signal-to-noise SEDs as well as images.

The time taken for each model to run varies with optical depth and covering factor of circumstellar material as seen from the radiation source. 
A disk model typically takes an hour to run on a 3 GHz Intel processor, and an embedded protostar can take 10 hours.
The total CPU time for the entire model grid, which was run on three different clusters, was approximately 65,000 hours (or roughly three weeks using approximately 60 CPUs).

The grid of models is available on a dedicated web server\,\footnote{http://www.astro.wisc.edu/protostars}.
This includes SEDs and polarization spectra for each inclination and aperture of each model.
Various components of the SED can be viewed separately, such as the flux emitted by the disk, star, or envelope, the scattered flux, and the direct stellar flux, as described in Section \ref{s:seds}.
Fluxes and magnitudes for common filter functions are also available.

\subsection{Sampling of the model parameters}
\label{s:params}

In the following section we present a detailed description of the parameter sampling.
It is not possible to explore parameter space in its entirety in a completely unbiased manner.
Therefore we have had to make arbitrary decisions concerning the ranges of parameter values.
The parameter ranges covered by the grid of models span those determined from observations and theories, and can be divided into three categories: the central source parameters (stellar mass, radius and temperature), the infalling envelope parameters (the envelope accretion rate, outer radius, inner radius, cavity opening angle and cavity density), and the disk parameters (disk mass, outer radius, inner radius, flaring power, and scaleheight).
Also included is a parameter describing the ambient density surrounding the YSO.

All the masses, mass accretion rates and densities for the disk and envelope parameters assume a gas-to-dust ratio of $100$.
Note that the results can be scaled to different gas-to-dust ratios since only the dust is taken into account in the radiation transfer.
For example, a disk with a total mass of $0.01$\,$\msun$ in a region where the gas-to-dust ratio is $100$ will produce the same SED as a disk with a total mass of $0.1$\,$\msun$ in a region where the gas-to-dust ratio is $1000$ (such as a low-metallicity galaxy or perhaps the outer Milky Way).
In addition to disk masses, the following parameter values should be rescaled accordingly in regions where the gas-to-dust ratio is not $100$: the envelope accretion rates, disk accretion rates, cavity densities and ambient densities.
Note that it is only necessary to re-scale the parameter values, and that it is not necessary to re-run the radiation transfer models

\subsubsection{The Stellar Parameters}

The parameters for the 20,000 YSO models were chosen using the following procedure. First a stellar mass $\mstar$ was randomly sampled from the following probability distribution function:

\begin{equation}
f(\mstar)d\mstar=\frac{1}{\log_{e}{10}}\frac{1}{\log_{10}{M_{2}}-\log_{10}{M_{1}}}\frac{d\mstar}{\mstar}
\end{equation}

The masses were sampled between $M_{1}=0.1M_{\odot}$ and $M_{2}=50M_{\odot}$. This produced a constant density of models in $\log_{10}{\mstar}$ space. Following this, a random stellar age $\agestar$ was sampled from a similar probability distribution function:

\begin{equation}
f(\agestar)d\agestar=\frac{1}{6}~\frac{1}{t_{\rm{max}}^{1/6}-t_{\rm{min}}^{1/6}}~\frac{d\agestar}{\agestar^{5/6}}
\end{equation}

\begin{figure}
\epsscale{0.38}
\plotone{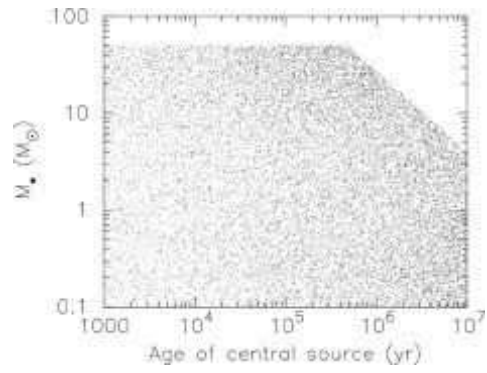}
\caption{The values of the central source mass $\mstar$ and evolutionary age $\agestar$ for the 20,000 models. Each point represents one model.\label{f:agemstar}}
\end{figure}

The ages were sampled between $t_{\rm{min}}=10^3$\,yr and $t_{\rm{max}}=10^7$\,yr. This distribution produced a density of models close to constant in $\log_{10}{\agestar}$ space, but with a slight bias towards larger values of $\agestar$. This was done to avoid a deficit of disk-only models. In cases where the resulting values of the stellar age were greater than the combined pre-main sequence and main-sequence lifetime of the star (estimated from the sampled masses $\mstar$), the age was resampled until it was within the adequate lifetime. The values of $\mstar$ and $\agestar$ for the 20,000 models are shown in Figure~\ref{f:agemstar}.

\begin{figure}
\epsscale{0.8}
\plotone{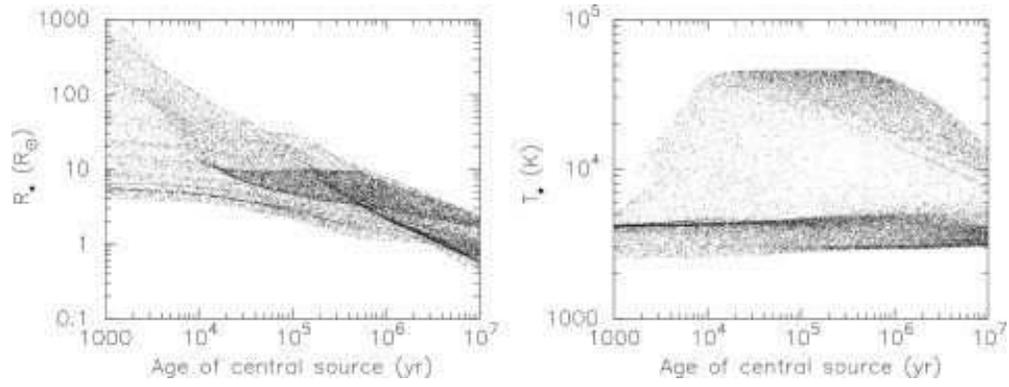}
\caption{The values of the central source radius $\rstar$ (left) and temperature $\tstar$ (right) as a function of evolutionary age for the 20,000 models. These values were derived from the central source mass and evolutionary age using evolutionary tracks.\label{f:rstartstar}}
\end{figure}

For each set of $\mstar$ and $\agestar$ the values of the stellar radius $\rstar$ and temperature $\tstar$ were found by interpolating pre-main sequence evolutionary tracks (\citealt{bernasconi96} for $\mstar\ge9\msun$; \citealt{siess00} for $\mstar\le7\msun$; a combination of both for $7\msun<\mstar<9\msun$).
The values of $\tstar$ and $\rstar$ are shown in Figure~\ref{f:rstartstar}.
It is important to note that the evolutionary age of the central sources is not a parameter in the radiation transfer code, and was only used to get a coherent radius and temperature as well as approximate ranges of disk and envelope parameters. 
Any errors in the evolutionary tracks can easily be accommodated by the large range of parameter values allowed at a given age.
Furthermore it is always possible to reassign a stellar source age to a model if it is found that a different set of evolutionary tracks is more appropriate for pre-main-sequence stars

Once the stellar parameters were determined, these were used to find the disk and envelope parameters.
Since there exists no exact relation between the parameters of the central star and those of the circumstellar environment, we sampled values of the various parameters from ranges that are functions of the evolutionary age of the central source, as well as functions of the stellar masses in certain cases.
These ranges were based on theoretical predictions and observations.

\subsubsection{The infalling envelope parameters}

\begin{figure}
\epsscale{0.8}
\plotone{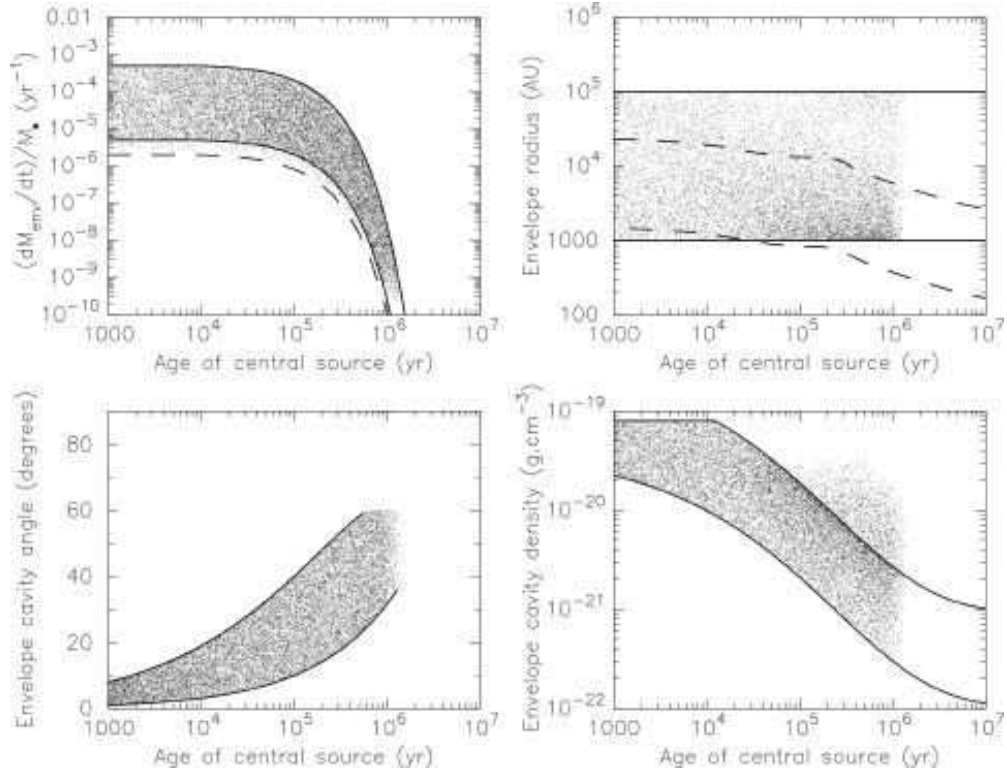}
\caption{The envelope parameters as a function of the evolutionary age for the 20,000 models. For the envelope accretion rate (top left), the solid lines show the range that was sampled from, for models with $\mstar<20\msun$. For models with $\mstar>20\msun$, we sampled from the same range of $\mdote$ as a $20\msun$ model, leading to lower values of $\mdote/\mstar$. The dashed line shows the lower limit of $\mdote/\mstar$ values for models with $\mstar>20\msun$. For the envelope radius (top right), the dashed lines represent the original range that was sampled from, for a $1\msun$ central source. For the opening angle of the bipolar cavities (bottom left), values above $60^\circ$ were reset to $90^\circ$. For the cavity density (bottom right), the values were sampled between the two solid lines; subsequently, any value of the cavity density that was smaller than the ambient density was reset to the ambient density.\label{f:envelope}}
\end{figure}

The (azimuthally symmetric) density structure $\rho(r,\theta)$ for the rotationally flattened infalling envelope is given in spherical polar coordinates by \citep{ulrich76,terebey84}
\begin{equation}
\rho(r,\theta) = \frac{\mdote}{4\pi \left(G \mstar \rc^3\right)^{1/2}}\left(\frac{r}{\rc}\right)^{-3/2}\left(1+\frac{\mu}{\mu_0}\right)^{-1/2}
\left(\frac{\mu}{\mu_0}+\frac{2\mu_02 \rc} {r}\right)^{-1},
\end{equation} 
where $\mdote$ is the envelope accretion rate, $\rc$ is the centrifugal radius, $\mu=\cos{\theta}$ ($\theta$ is the polar angle), and $\mu_0$ is cosine of the angle of a streamline of infalling particles as $r\rightarrow\infty$.
The centrifugal radius $\rc$ determines the approximate disk radius and flattening in the envelope structure. We solve for $\mu_0$ from the equation for the streamline:
\begin{equation} 
\mu_0^3 +\mu_0(r/\rc-1)-\mu(r/\rc)=0.
\end{equation}
The range of values sampled for each of the envelope parameters, as well as the final parameter values, are shown in Figure~\ref{f:envelope}.

\paragraph{Envelope accretion rate} We sampled values of $\mdote/\mstar$ from an envelope function that is constant for $\agestar<10^4$\,yr \citep*{shu77,terebey84}, decreases between $10^5$\,yr and $10^6$\,yr \citep{foster93,foster94,hartmann01} and goes to zero around $10^6$\,yr \citep{young05_2}. The range of accretion rate values $\mdote/\mstar$ at a given time is two orders of magnitude wide, and the values were sampled uniformly in $\log{\mdote/\mstar}$.  The average values were chosen to match estimates of low-mass \citep{adams87,kenyon93p1,kenyon93p2,whitney97} and high-mass YSOs \citep{wolfire87,osorio99,omukai01,churchwell02,yorke02,mckee03,bonnell04}. In cases where this value fell below $10^{-9}\left(\mstar/M_{\odot}\right)^{1/2}\rm{yr}^{-1}$, the envelope was discarded, and the YSO model was considered as a disk-only model. Finally, for stellar masses above $20$\,$\msun$ the envelope accretion rate was sampled from the same range of $\mdote$ as a $20$\,$\msun$ model, so that the largest value of  $\mdote/\mstar$ is $5\times10^{-4}$\,yr$^{-1}$.

\paragraph{Envelope outer radius} To find the envelope outer radius, we first calculated the approximate radius at which the optically thin radiative equilibrium temperature falls to $30$\,K \citep{lamers99}:
\begin{equation}
R_{0}=\frac{1}{2}\rstar\left(\frac{\tstar}{30\,{\rm K}}\right)^{2.5}.
\end{equation}
We then sampled a random value for $\rmaxe$ uniformly in $\log{R}$ space between $R_{0}\times 4$ and $R_{0}/4$ (the latter to account for truncation by tidal forces in clusters).
The initial range of values for a $1$\,$\msun$ star is shown as an example in Figure~\ref{f:envelope}.
In cases where $\rmaxe>10^5$\,AU, $\rmaxe$ was set to $10^5$\,AU.
In cases where $\rmaxe<10^3$\,AU and $R_{0}\times4>10^3$\,AU, we re-sampled $\rmaxe$ between $10^3$\,AU and $R_{0}\times4$.
Finally, in cases where $R_{0}\times4<10^3$\,AU, $\rmaxe$ was set to $10^3$\,AU.

\paragraph{Envelope cavity opening angle} In the current grid of models we use a cavity shape described in cylindrical polar coordinates by $z=c\varpi^d$ where $\varpi$ is the radial coordinate. 
For all of our models, we fix $d=1.5$, and taking the cavity opening angle $\thetacav$ to be that for which z=$\rmaxe$, $c$ is given by $\rmaxe/(\rmaxe\tan{\thetacav})$.
The cavity opening angle was sampled from a range of values increasing with evolutionary age, as indicated by observations of cavities and outflows in Class~0 and Class~I protostars \citep{zealey93,chandler96,tamura96,lucas97,hogerheijde98,velusamy98,bachiller99,padgett99,beuther05,shepherd_iau,arce01,arce04,arce06,arce06_ppv,ybarra06}.
In cases where the cavity angle was larger than $60^{\circ}$, it was reset to $90^{\circ}$, assuming that the envelope is mostly dispersed at this stage of evolution.

\paragraph{Envelope cavity density} The density of gas and dust in the envelope cavity was sampled from a range of values following a decreasing function of time, one order of magnitude wide, with values ranging between $10^{-22}$ and $8\times10^{-20}$\,g\,cm$^{-3}$. These values correspond to molecular number densities of $n_{\rm H_2}=3\times10^1\rightarrow2\times10^4$\,cm$^{-3}$ that are typical of observations of molecular outflows \citep[e.g.][]{moriarty95,moriarty95_p3}. In cases where the cavity density was lower than the ambient density of the surrounding medium (described in the next paragraph) the cavity density was reset to the ambient density. The cavity density is assumed to be constant with radius from the central source for simplicity (as one would expect from a cylindrical outflow with a constant outflow rate).

\paragraph{The ambient density} The infalling envelope is assumed to be embedded in a constant density ISM.
This surrounding ambient density can contribute to the extinction, scattering
and thermal emission of the circumstellar dust, especially in high-luminosity
sources, which can heat up large volumes of the surrounding molecular cloud.
In disk-only sources and low-mass YSOs, the contribution from the ambient
density is small, though often non-zero.   This can be seen 
on our website in disk-only models by selecting the contribution from the envelope
(in this case, the ambient density)
to the SED.
We sampled this density between $1.67\times10^{-22}(\mstar/\msun)$\,g\,cm$^{-3}$ (or $10^{-22}$\,g\,cm$^{-3}$, whichever was largest) and $6.68\times10^{-22}(\mstar/\msun)$\,g\,cm$^{-3}$, corresponding to values in the range $n_{\hh}\sim50-200\,(\mstar/\msun)$\,cm$^{-3}$.
This is consistent with typical densities of $n_{\hh}\sim100$\,cm$^{-3}$ observed in molecular clouds \cite[e.g.][]{blitz93}.

\subsubsection{The disk parameters}

For the disk structure, we use a standard flared (azimuthally symmetric) accretion disk density \citep{shakura73,lyndenbell74,pringle81,bjorkman97,hartmann98}:

\begin{equation}
\rho(\varpi,z)=\rho_0 
\left[1-\sqrt{\frac{R_\star}{\varpi}}\right]
\left(\frac{R_\star}{\varpi}\right)^\alpha
\exp{\left\{{ -\frac{1}{2}\left[\frac{z}{h}\right]^2}\right\}}\; ,
\end{equation}

\noindent
where $h$ is the disk scaleheight, which increases with radius as $h\propto\varpi^\beta$, $\beta$ is the flaring power, and $\alpha=\beta+1$ is the radial density exponent.
The disk scaleheight at the dust sublimation radius is set to be that for hydrostatic equilibrium, multiplied by a factor $\zmin$.
This factor can be smaller than one if there is gas or other opacity inside the dust destruction radius, decreasing the amount of stellar flux incident on the inner wall; or it can be used to mimic dust settling, as described in the disk structure paragraph in this section.
The normalization constant $\rho_0$ is defined such that the integral of the density $\rho(\varpi,z)$ over the whole disk is equal to $\mdisk$.
The values for all the disk parameters are shown in Figure~\ref{f:disk}.

\paragraph{Disk mass} The disk mass was sampled from $\mdisk/\mstar\sim0.001-0.1$ at early evolutionary stages, and a wider range of masses between $1$ and $10$\,Myr. This allows for disk masses of $\sim0.001-0.1\msun$ typically observed in low-mass YSOs, whether during the early infall phase or during the T-Tauri phase \citep[e.g.][]{beckwith90,terebey93,dutrey96,kitamura02,looney03,andrews05}, higher disk masses around high-mass YSOs, and disk masses down to $\mdisk/\mstar=10^{-8}$ after $1$\,Myr to allow for the disk dispersal stage.

\paragraph{Disk outer radius and envelope centrifugal radius} In a rotating and infalling envelope, material near the poles has little angular momentum, and will fall near to or onto the central source.
In contrast, infalling material close to or along the equatorial plane will have the most angular momentum, and will fall to a radius $\rc$ in the equatorial plane, where $\rc$ is the centrifugal radius \citep[e.g.][]{cassen81,terebey84}.
Therefore, the centrifugal radius is usually associated with the outer radius of the circumstellar disk.
We sampled values of $\rc$ between $1$\,AU and $10,000$\,AU, following a time-dependent range of values that allows for smaller radii in younger models.
Theories indicate that the centrifugal radius will grow with time, if the infalling envelope was initially rotating as a solid body \citep[e.g.][]{adams86}.
In Taurus, disk sizes can be imaged directly \citep{burrows96,padgett99} and are typically a few hundred AU for both Class~I and II sources.
Large disks have been indicated around high-mass stars (see recent review article by \citealt{cesaroni06} for an exhaustive list); however, some of these observations may have been detecting the envelope toroid, which is typically twice as big as the centrifugal radius in our models.
Most observations of massive YSOs suggest typical disk sizes of roughly $500$ to $2,000$\,AU.

For disk-only models, the disk outer radii were sampled from the same time-dependent range of values as that used for the centrifugal radius for models with infalling envelopes.
Subsequently, two thirds of disk-only models saw their outer radius truncated.
To do this, we reset the disk outer radius to be randomly sampled between $10$\,AU and the centrifugal radius.
This was done to account for the possible truncation of the outer regions of disks in dense clusters due to stellar encounters or photoevaporation (see e.g. observations by \citealt{vicente05} who find that only $\sim50$\,\% of disks in the Trapezium cluster are larger than $50$\,AU; see also simulations of cluster formation by \citealt{bate03}).
The disk masses for these models were then recalculated by assuming the original density structure, and removing the truncated mass (not recalculating the disk mass could lead to unrealistically dense disks with small outer radii).

\paragraph{Disk (and envelope) inner radius} For all models, the envelope inner radius was set to the disk inner radius. For one third of all our models, the disk inner radius was set to the dust destruction radius $R_{\rm sub}$ empirically determined to be \citep{whitney04}:
\begin{equation}
R_{\rm sub}=\rstar(T_{\rm sub}/\tstar)^{-2.1}
\end{equation}
where we adopt $T_{\rm sub}=1,600$\,K as the dust sublimation temperature. In the remaining two thirds of models, the inner disk/envelope radii were increased. This was done in order to account for binary stars clearing cavities in envelopes in young sources \citep[e.g.][]{jorgensen05}, and binary stars or planets clearing out the inner disk in disk models \citep{lin79_1,lin79_2,artymowicz94,calvet02,rice03}. For these models, the disk and envelope inner radius were sampled between the dust destruction radius and $100$\,AU (or the disk outer radius in cases where this was less than $100$\,AU). The disk masses for these models were then recalculated (as for the disk outer radius).

\paragraph{Disk structure} The disk flaring parameter $\beta$ and the scaleheight factor $\zmin$ were sampled from a range that depends on the disk outer radius.
For large disks, the average $\beta$ decreases to prevent geometries that resemble envelopes with curved cavities (which would confuse the interpretation of the SEDs).
Both parameters $\beta$ and $\zmin$ together can be used to mimic the effects of dust settling (i.e. low values of $\beta$ and $\zmin$ can be an indication that dust has settled towards the disk midplane), and $\beta$ can also be used to describe various degrees of disk flaring \citep{kenyon87,miyake95,chiang97,dalessio98,furlan05,dalessio06}.

\paragraph{Disk accretion rate} In order to sample different disk accretion rates, we sampled values of the disk $\alpha_{\rm disk}$ parameter \citep{shakura73,pringle81} logarithmically between $10^{-3}$ and $10^{-1}$.
The disk accretion rate is calculated in the radiation transfer code using \citep{pringle81,bjorkman97}:
\begin{equation}
\label{e:mdotdisk}
\dot\mdisk= \sqrt{18\pi^3}\,\alpha_{\rm disk}\,V_c\,\rho_0\,h_0^3/\rstar\;,
\end{equation}
with the critical velocity $V_c=\sqrt{G\mstar/\rstar}$.
Accretion shock luminosity on the stellar surface is included following the method of \citet{calvet98}.
Average disk accretion values are based on the literature \citep[e.g.][]{valenti93,hartigan95,hartmann98,calvet04}.

\subsubsection{Caveats for the current set of models}
\label{s:caveats}

We now point out approximations of our model grid that will be addressed in future versions.
\begin{itemize}
\item Accretion from the envelope is not accounted for as a luminosity source (as distinct from disk accretion luminosity, which is included). This is likely only important in the very youngest sources, which may not be detected at IRAC wavelengths (our primary modeling goal at present). In the inside-out collapse model \citep{shu87}, infall occurs from further out in the envelope as time proceeds. In a rotating envelope, this material has more angular momentum and thus impacts further out in the disk, liberating smaller amounts of accretion energy.  
Once enough mass builds up in the disk to cause it to be gravitationally unstable, large accretion events may occur through the disk \citep{kenyon90,hartmann93,hartmann96,vorobyov05,green06}.  Therefore, if this scenario applies in most sources, the stellar and disk accretion luminosity included in our models should be adequate.
\item The models in this grid do not include multiple source emission (although the radiation transfer code has this capability).
We partially account for this by allowing for large inner envelope and disk holes at all evolutionary stages, which is probably the main effect of a binary system. Whether one or two sources illuminates an envelope from the inside likely has little effect on the emergent SED, but the size of the inner hole created by a binary star system has a large effect on the SED \citep{jorgensen05}. 
Currently, our inner holes are completely evacuated.
Future versions of the grid will have partially evacuated inner holes that could affect the mid-IR SED.
\item The envelope geometry is assumed to be dominated by free-fall rotational collapse.
This is a good approximation in the inner regions because rotation and free-fall likely dominate over magnetic effects \citep{galli93,desch01,nakamura99} (though see \citet{allen03} for a discussion of magnetic braking), and observations of radial intensity profiles are consistent with free-fall collapse \citep{chandler00,vandertak00,beuther02,mueller02,young03,hatchell03}.
In the outer regions, magnetic fields, turbulence, and other non-ideal initial conditions could affect the density distributions \citep{foster93,bacmann00,whitworth01,allen03,goodwin04,galli05}.
However, the mid-IR fluxes are most sensitive to the inner envelope, bipolar cavities, and disk geometries.
\item The shape of the outflow cavities, and density distributions are uncertain. We are working with theorists and observers to improve our understanding of these.
\item The current grid of models does not include heating by the external interstellar radiation field. This is an important heating source for very low-luminosity sources \citep{young04}, and was recently shown to be important for the temperature structure and chemistry of high-mass protostellar envelopes \citep{jorgensen06}.
\item The current grid of models does not include brown dwarfs or brown dwarf precursors.
\item Our dust models do not include ice coatings.  In addition, our debris disk dust models
are probably not appropriate.  These will be improved in the next grid of models,
in collaboration with experts in these areas \citep[e.g.][]{ossenkopf94,wolf03,wolf04,carpenter05}.
\item The code does not account for PAH or small-grain continuum emission, which can contribute to mid-IR emission in YSOs with hot central stellar sources \citep*{habart04,ressler03}.
\item The flared disk geometries used here may not be appropriate for the very high-mass sources where photoionization can drive a wind and essentially puff up the disk \citep{hollenbach94,lugo04}.
\item The stellar evolutionary tracks that we use are for canonical non-accreting pre-main-sequence stars \citep{bernasconi96,siess00}.
As mentioned in Section~\ref{s:params}, the evolutionary tracks are not crucial in the sense that they are only used to get approximate values of consistent stellar radius and temperature for a star of a given mass and approximate evolutionary `age'.
\item Massive, luminous stars with large inner dust holes (due to the large dust destruction radius) may have optically thick {\it gas} inside the dust hole, which we do not account for. This would add more near-IR flux from the hotter gas.
\item A different geometry may be necessary for very massive dense stellar clusters; that is, we should include a cluster of stars embedded in an envelope rather than one star \citep{zinnecker93,hillenbrand97,clarke00,sollins05,allen_ppv}.
\item Examination of our grid of models shows a jump in A$_V$ in high-mass sources
between those without envelopes and those with even low-density envelopes.  This
is due to the fact that we set a floor to the envelope density at the ambient density,
and the outer radii of envelopes around high-luminosity sources are large.  This
was intentional to account for the fact that high-luminosity sources heat up large
volumes of the surrounding molecular cloud.  However, we did not consider the fact that
high-mass sources can also disperse material with their stellar winds.  Our future grids
will allow the ambient density to go lower to account for effects such as wind-blown
cavities and dispersed interstellar medium.
\item Since the emergent flux from the Monte Carlo code is binned into direction (in equal intervals of
$\cos \theta$), it is effectively averaged over a finite range in angle.  If the SED changes rapidly with angle, for example, in a geometrically thin edge-on disk source viewed near edge-on, these effects will be washed out.  Even though the edge-on angle bin has a fairly narrow range (from 87-93$^{\circ}$), more flux emerges from 87$^{\circ}$ and 93$^{\circ}$ than from 90$^{\circ}$, and the ``edge-on'' SED, centered on $\theta=90^{\circ}$, will reflect that of a slightly less edge-on source. 
In future grids, we will calculate SEDs at specific outgoing angles, removing this averaging effect. 
\end{itemize}

We plan to address all of these issues in future versions of the grid of models.
We also hope to get suggestions for other improvements from theorists and observers alike.
However, we believe the current model grid should be adequate to model a large range of stellar masses and evolutionary stages, with the exception of very low-luminosity sources ($L<0.2L_\odot$) and sources in very dense clusters ($n>1,000$~stars\,pc$^{-3}$).

\section{Results and Analysis}
\label{s:results}

\subsection{Evolutionary stages}
\label{s:stage}

As mentioned in Section~\ref{s:intro}, YSOs are traditionally grouped into three `Classes' according to the spectral index $\alpha$ of their SED, typically measured in the range $\sim2.0-25$\microns \citep{lada87}. 
Additionally, Class~0 objects are taken to refer to sources that display an SED similar to a 30K graybody at sub-mm wavelengths, with little or no near- and mid-IR flux \citep*{andre93}.

The spectral index classification (`Class') can sometimes lead to confusion in terminology, as it has in many cases become synonymous with evolutionary stage; yet, the same object can be classified in different ways depending on viewing angle.
For example, an edge-on disk can have a positive spectral index that would make it a Class~I object; and a pole-on ``embedded'' source might have a flat spectrum, rather than a rising one \citep{calvet94}.
Other effects, such as increased stellar temperature (above the canonical Taurus value of $4,000$\,K) can lead to positive spectral indices in disk sources \citep{whitney04}.

In discussing the evolutionary stages of our models, we adopt a `Stage' classification analogous to the `Class' scheme, but referring to the actual evolutionary stage of the object, based on its physical properties (e.g. disk mass or envelope accretion rate) rather than properties of its SED (e.g. slope).
Stage~0 and I objects have significant infalling envelopes and possibly disks, Stage~II objects have optically thick disks (and possible remains of a tenuous infalling envelope), and Stage~III objects have optically thin disks. 

Using this classification alongside the spectral index classification can help avoid any confusion between observable and physical properties:
for example, we would refer to an edge-on disk as a Stage~II object that may display a Class~I SED (rather than an `edge-on Class~II' object).

The exact boundaries between the different Stages are of course arbitrary in the same way as the Class scheme.
In the following sections, we choose to define Stage~0/I objects as those that have $\mdote/\mstar>10^{-6}$\,yr$^{-1}$, Stage~II objects as those that have $\mdote/\mstar<10^{-6}$\,yr$^{-1}$ and $\mdisk/\mstar>10^{-6}$, and Stage~III objects as those that have $\mdote/\mstar<10^{-6}$\,yr$^{-1}$ and $\mdisk/\mstar<10^{-6}$.  Note that we have grouped Stage~0 and I
objects together, and refer to them throughout this paper as Stage~I objects. 
 
\subsection{SEDs and Polarizations of Selected Models}
\label{s:seds}

Figure~\ref{f:seds} shows example SEDs from our grid of models. We show SEDs of low, intermediate, and high mass stars at three stages of evolution.
Each panel shows SEDs for the ten viewing angles, with the top spectrum corresponding to the pole-on viewing angle, and the bottom spectrum corresponding to the edge-on viewing angle.
Also shown for each SED is the input stellar photosphere.
Longward of $10$\micronsns, the flux is due mainly to reprocessing of absorbed stellar flux by the circumstellar dust.
Shortward of $10$\microns the flux is dominated by stellar light (e.g. in Stage~III or pole-on Stage~I \& II models), scattered stellar light (e.g. edge-on Stage~I \& II models), and warm dust from the inner disk and the bipolar cavities \citep{whitney04}.

Since we are using a Monte-Carlo radiation transfer code, it is possible to track various properties as photons propagate through the grid, and for instance to flag each one by its last point of origin.
For example, an energy packet absorbed and re-emitted by the disk is considered to have a point of origin in the disk.
Scattering is not considered a point of origin; thus, a photon re-emitted by the disk that scatters in the envelope is considered a disk photon.
Figure~\ref{f:seds_indiv} shows SEDs for the same models as before, making use of this information.
We show three separate SEDs for each model and inclination: the blue SED shows the energy packets whose last point of origin is in the stellar photosphere, the green SED shows the energy packets whose last point of origin is in the disk, and the red SED shows the energy packets whose last point of origin is in the envelope.
Energy packets originating directly from the star contribute a significant amount of emission in pole-on Stage~I objects as well as in Stage~II and III objects.
Energy packets last emitted by the envelope clearly dominate the SEDs of Stage~I models longwards of $10$\micronsns.
Finally, energy packets last emitted by the disk dominate the SEDs of Stage~II objects longward of near-IR wavelengths.
Note that the envelope emission in disk-only Stage~II and III sources is due to the ambient density.
Other photon properties tracked in the radiation transfer include whether a given photon was last scattered before escaping, or whether a photon escaped directly from the stellar surface without interacting with the circumstellar dust.
All these SED components are available for each model on our web server.

Figure~\ref{f:seds_pol} shows polarization results for the same models as Figures~\ref{f:seds} and \ref{f:seds_indiv}.
The current grid of models does not have sufficient signal-to-noise to produce high-resolution polarization spectra.
However, the polarization can be smoothed and convolved with broadband filter profiles.
Because these models are axisymmetric and include scattering from spherical grains (no dichroism from aligned grains), the polarization can be fully described by the $Q$ stokes parameter, $Q = P \cos\chi$, where $P$ is the degree of linear polarization, and $\chi$ is the orientation of the polarization with respect to the axis of symmetry (in these models, this is perpendicular to the disk plane, or parallel to the presumed outflow axis).
Negative $Q$ polarization indicates that the polarization is aligned parallel to the disk plane, and positive $Q$ polarization is aligned perpendicular to the disk plane.  
The first thing to note about the results in Figure~\ref{f:seds_pol} is that the polarization sign varies with wavelength and Stage.  This is shown more clearly in Figure~\ref{f:k_pol}, which shows K-band polarization from the entire grid of models as a function of evolutionary stage for four selected inclinations.
The highly embedded Stage~I sources have negative polarization (that is, aligned parallel to the disk plane); and the less embedded Stage~I and Stage~II sources have the opposite sign.
This is an optical depth effect, as discussed in \citet{bastien87}, \citet{kenyon93p2} and \citet{whitney97}.
As an optical depth effect, it is also a wavelength effect, since the dust opacity decreases with increasing wavelength.
As shown in Figure 9, the high-mass Stage~I source exhibits a sign change with wavelength.
This is a clear indication of a Stage~I source, since a Stage~II source always has polarization oriented perpendicular to the disk plane.
A Stage I source has polarization aligned parallel to the disk at short wavelengths, where scattering in the outflow cavities is the main source of polarization, and the disk is too deeply embedded to be visible.
At longer wavelengths the envelope becomes more optically thin, and scattering in the disk plane dominates the polarization, causing it to become aligned perpendicular to the disk.
If a sign change in the polarization is observed, this information can be used to distinguish a Stage~I source from an edge-on Stage~II source, which could have a similar SED.
However, we note that the reverse is not necessarily true: the absence of a sign change in the polarization does not necessarily rule out that the source is a Stage~I object, since a sign change can occur outside the observed wavelength range.

\subsection{Spectral index classification}
\label{s:si}

\subsubsection{The dependence of spectral index on evolutionary stage}
\label{s:sivsparams}

In this section, we examine the spectral indices of our model SEDs.
The traditional definition of the spectral index of an SED \citep{lada87} is its slope in $\log_{10}{\lambda F_\lambda}$ vs $\log_{10}{\lambda}$ space, longward of 2\micronsns.

This can be taken as the slope in $\log_{10}{\lambda F_\lambda}$ versus $\log_{10}{\lambda}$ space of the line joining two flux measurements at wavelengths $\lambda_1$ and $\lambda_2$, or as the slope of the least-squares fit line to \emph{all} the flux measurements between and including the two wavelengths $\lambda_1$ and $\lambda_2$.
We use the notation \alphaend to refer to the former, and $\alpha_{[\lambda_1\rightarrow\lambda_2]}$ to refer to the latter of these two definitions.
Both definitions have been used in the literature (e.g. \citealt{myers87} and \citealt{kh95} use \alphaendns, while \citealt{haisch01}, \citealt{lada06}, and \citealt{jorgensen06} use \alphaallns), and in some cases it is not explicitely stated which choice has been made.
In this section, we choose to use the \alphaend definition of spectral index, as it is unique for each SED for a given $\lambda_1$ and $\lambda_2$, whereas \alphaall depends on which points are included inside the range.

As noted previously (Section \ref{s:output}), each SED is computed in 50 different apertures.
The main effect of varying the aperture is a bluing of the colors of Stage~I sources with larger aperture, as discussed in \citet{whitney03p1}.
To match typical point-source photometry observations, we use our $2760$\,AU aperture, which corresponds to $\sim3$'' at a distance of 1\,kpc.

Figure~\ref{f:sivsparams} shows the dependence of three different spectral indices on the envelope accretion rate and the disk mass.
The following spectral indices were computed for all the model SEDs in the grid: $\alpha_{[\rm{IRAC}\,3.6\,\&\,8.0]}$, $\alpha_{[J\,\&\,\rm{IRAC}\,8.0]}$, and \alpharef.
In this section and following sections we refer to \alpharef as our `reference' spectral index since it is close to the commonly used $2.0-25$\microns range.

In order to calculate a spectral index using either method from a given set of fluxes, we require these fluxes to have a signal-to-noise of at least 2.
The hashed box in Figure~\ref{f:sivsparams} shows the values of $\mdote/\mstar$ for which at least $10$\,\% of SEDs had insufficient signal-to-noise for the spectral index calculation.

We first look at how the reference spectral index (shown in the bottom panel of Figure~\ref{f:sivsparams}) depends on the disk mass and the envelope accretion rate:
\begin{itemize}
\item The tendency is for the spectral index to increase as the disk mass and the envelope accretion rate increase.
Therefore, on average, younger sources tend to have a larger spectral index.
However, for a given disk mass or envelope accretion rate, the spread in the values of the spectral index is important.
This suggests that although in large samples of sources larger spectral indices are likely to indicate youth, the spectral index of an individual source is not a reliable indicator of its evolutionary stage.
For example, a source with a reference spectral index of $-0.5$ could have virtually any disk mass or envelope accretion rate.
\item For disk masses $\mdisk/\mstar<10^{-5}$, the spectral index increases proportionally to $\log_{10}{\mdisk/\mstar}$, whereas above this limit the spectral index is independent of disk mass.
This is because for low disk masses the disk is optically thin, and near- and mid-IR radiation is seen from the whole disk, whereas for larger disk masses the disk is optically thick, and the near- and mid-IR radiation is only seen from the surface layers of the disk.
This is in agreement with the results from \citet{wood02}.
\item For envelope accretion rates $\mdote/\mstar<10^{-6}$\,yr$^{-1}$, the spectral index does not vary with $\mdote/\mstar$.
This indicates that the envelope is optically thin and does not contribute significantly to the SED, i.e. the SED is dominated by disk or stellar emission.
\item For envelope accretion rates $\mdote/\mstar>10^{-6}$\,yr$^{-1}$ the range of possible spectral indices widens and the upper limit of this range increases roughly linearly with $\log_{10}{\mdote/\mstar}$.
The widening of the range is likely due to the high dependence of Stage~I colors on viewing angle \citep{whitney03p1} and stellar temperature \citep{whitney04}.
The increase of the upper limit is expected: as the accretion rate increases the envelope becomes more optically thick, and progressively obscures the central source and the regions of high-temperature dust, reddening the SED.
\item The decrease in the upper range of spectral indices for $\mdote/\mstar>5\times10^{-5}$\,yr$^{-1}$ is an artifact due to the signal-to-noise requirements for computing a spectral index: models with heavily embedded sources have very few or no energy packets emerging at near-IR wavelengths, leading to a poor signal-to-noise at these wavelengths.
The majority of models with high accretion rates for which the reference spectral index can be calculated are pole-on or close to pole-on.
For these SEDs the star is not heavily obscured by the envelope since one is looking down the cavity, and the spectral index is bluer.
For models with $\mdote/\mstar>10^{-4}$\,yr$^{-1}$, $98$\,\% of SEDs for the pole-on viewing angle have sufficient signal-to-noise at near- and mid-IR, as opposed to only $23$\,\% of SEDs for the edge-on viewing angle (this should improve in future grids of models, as we plan to produce higher signal-to-noise models).
\item Our choice of boundary values for the `Stage' classification seems to be appropriate, since most Stage~I models have $\alpha>0$, most Stage~II models have $-2<\alpha<0$, and a large fraction of Stage~III models have $\alpha<-2$.
This means that for the majority of models, the `Stage' is equivalent to the `Class'.
\end{itemize}

The two top panels of Figure~\ref{f:sivsparams} show the spectral indices computed over a narrower wavelength range ($\alpha_{[\rm{IRAC}\,3.6\,\&\,8.0]}$ and $\alpha_{[J\,\&\,\rm{IRAC}\,8.0]}$). These show a similar pattern, albeit the spread in spectral indices is much larger for a given disk mass or envelope accretion rate.
This shows that including a longer wavelength flux measurement ($\sim20$\micronsns) in the spectral index calculation provides a better indicator of evolutionary stage.

\subsubsection{The dependence of the spectral index of disk-only sources on stellar temperature, disk inner radius, and disk flaring power}

\label{s:sivstemp}

As mentioned above, the spectral indices of the SEDs for disk-only models with $\mdisk/\mstar>10^{-5}$ are independent of $\mdisk/\mstar$.
For these models, the spread in spectral indices for a given $\mdisk/\mstar$ is due mainly to the spread in stellar temperatures, disk inner radii, and disk flaring powers; this is illustrated in Figure~\ref{f:siadditional}, which shows the dependence of the spectral index of these SEDs on the stellar temperature $\tstar$, the disk inner radius $\rmind$, and the disk flaring power $\beta$.

The left-hand panels shows all disk-only models with $\mdisk/\mstar>10^{-5}$.
For low temperatures ($\sim3,000-5,000$\,K), the $\alpha_{[\rm{IRAC}\,3.6\,\&\,8.0]}$ and the $\alpha_{[J\,\&\,\rm{IRAC}\,8.0]}$ spectral indices are separated into two distinct groups.
The largest group, centered between spectral index values of $-2$ and $-1$ is the bulk of the disk models with no large inner holes.
The smaller group, centered at lower spectral index values, represent the models which include inner holes large enough that the JHK and IRAC fluxes are purely photospheric.
We note that in the case of the $\alpha_{[J\,\&\,\rm{IRAC}\,8.0]}$ spectral index, and to a lesser extent the $\alpha_{[\rm{IRAC}\,3.6\,\&\,8.0]}$ spectral index, the average value of these two groups decreases as the temperature increases from $3,000$\,K to $5,000$\,K.
This is expected, as the colors of stellar photospheres at near-IR wavelengths for these temperatures are still dependent on the stellar temperature, and become bluer for larger temperatures.

Beyond $5,000$\,K, the spectral indices increase with stellar temperature, at least in the case of the $\alpha_{[\rm{IRAC}\,3.6\,\&\,8.0]}$ and the $\alpha_{[J\,\&\,\rm{IRAC}\,8.0]}$ spectral indices.
This is due to a lower contribution of the stellar flux compared to the infrared dust spectrum at these wavelengths \citep{whitney04}.
For models with a $5,000$\,K central source, this central source contributes significantly to the emergent spectrum, making the emission less red.
For a $10,000$\,K stellar source, the relative fraction of stellar flux to emission from the disk at near- and mid-IR wavelengths is much lower, leading to a more pure dust spectrum that is red at near- and mid-IR wavelengths.
This can be seen in the SEDs for the Stage~II and III models in Figures~\ref{f:seds} and \ref{f:seds_indiv}: although the general shape of the contribution to the SED from the disk does not change significantly between the low-, intermediate-, and high-mass models, the change in the stellar spectrum leads to redder colors at near- and mid-IR wavelengths for the high-mass (and therefore higher temperature) model.

The apparent gap in the models between $6,000$\,K and $10,000$\,K is due to the sampling of the model parameters using evolutionary tracks.
For $\alpha_{[\rm{IRAC}\,3.6\,\&\,8.0]}$ and the $\alpha_{[J\,\&\,\rm{IRAC}\,8.0]}$, the average spectral index values are different on either side of this gap.
This leads to the bimodal distribution seen in Figure~\ref{f:sivsparams} for these spectral indices.

The central panels in Figure~\ref{f:siadditional} show the detailed variations of the spectral index with disk inner radius for all the disk-only models with $\mdisk/\mstar>10^{-5}$, and with stellar temperatures below $6,000$\,K.
To avoid overcrowding of the plot, we do not show models with $\rmind=1\,\rsub$.
In all three cases, the spectral index first increases, then decreases to reach photospheric levels.
The initial increase is due to removal of the hottest dust from the inner disk and redistribution of the SED to slightly longer wavelengths.
As the inner radius increases further, the amount of circumstellar material emitting in mid-IR wavelengths is reduced, and the mid-IR emission decreases.

Finally, in the right-hand panels of Figure~\ref{f:siadditional}, we show all the disk-only models with $\mdisk/\mstar>10^{-5}$, with stellar temperatures below $6,000$\,K and with $\rmind=1\,\rsub$, versus the disk flaring power.
A larger flaring power in a disk leads to an increasing surface intercepting the starlight, and therefore an increase in reprocessed radiation.
This has only a slight effect on the $\alpha_{[\rm{IRAC}\,3.6\,\&\,8.0]}$ and the $\alpha_{[J\,\&\,\rm{IRAC}\,8.0]}$ spectral indices, but has a more pronounced effect on the reference spectral index \alpharefns.

\subsubsection{The dependence of spectral index on wavelength range and distribution}

The wavelength range of the fluxes used to calculate the spectral index of a source is dependent on the observations available for that given source.
In most cases, the choice of this range is likely to affect the value of the spectral index itself.
For example, as seen in Section~\ref{s:sivsparams}, a spectral index calculated using K and \mips fluxes will in most cases differ from a spectral index calculated using IRAC 3.6\microns and IRAC 8.0\microns fluxes.

Figure~\ref{f:siranges} shows the correlation between six different \alphaend spectral indices (calculated using various combinations of JHK, IRAC, and MIPS broadband fluxes) and the reference spectral index \alpharefns.
As can be seen, the value of the spectral index is highly dependent on the range of data used.
In some cases the difference can be larger than 1.0, which could lead to a different `Class' being assigned to a source depending on what spectral index is used.

The spectral index of a source may also be sensitive to whether it is calculated using only fluxes at two wavelengths (i.e. \alphaendns), or whether it is calculated using all fluxes in a given wavelength range (\alphaallns).
Figure~\ref{f:sidensity} shows the correlation between four different \alphaend spectral indices (using various combinations of JHK, IRAC, and \mips broadband fluxes) versus the equivalent \alphaall spectral indices.

The difference between $\alpha_{[\rm{IRAC}\,3.6\,\rightarrow\,8.0]}$ and $\alpha_{[\rm{IRAC}\,3.6\,\&\,8.0]}$ is small, typically of the order of 0.1 or less, which is expected, as the wavelength range is fairly narrow and the SED should be close to a straight line. The other three spectral indices, \alpharefns, $\alpha_{[J\,\&\,\rm{IRAC} 8.0]}$, and $\alpha_{[J\,\&\,\rm{MIPS}\,24]}$, show a larger difference: in some cases, the spectral index calculated using the two methods differs by up to 0.5. 

\subsection{Color-color classification}
\label{s:color}

\subsubsection{Virtual Clusters}

The parameters of our grid of models were sampled in order to cover many stages of evolution and stellar masses.
However, the distribution of models in parameter space is not meant to be representative of a typical star formation region, since it is meant to encompass outliers as well as typical objects, and since the models are sampled uniformly or close to uniformly in $\log_{10}{\mstar}$ and $\log_{10}{\agestar}$.

Therefore, in the current section, as well as presenting color-color diagrams for the whole grid of models, we also present results using a subset of models.
This subset, which we will refer to as our \emph{virtual cluster} was constructed by re-sampling models from the original grid in order to produce a standard IMF for the stellar masses \citep{kroupa01}, and to produce a distribution of ages distributed linearly in time, rather than logarithmically.
We decided to sample the stellar masses between $0.1$ and $30\msun$, and ages between $10^3$ and $2\times10^6$\,yr.
In this way we aim to reproduce the distribution of stellar masses and ages that might be expected from a cluster with the chosen IMF and a continuous star formation rate having switched on $2\times10^6$\,yr ago.
The distribution of masses and ages for the original grid of models and the virtual cluster are shown in Figure~\ref{f:cluster}.

We assumed two distances to this cluster: $250$\,pc to mimic a relatively nearby star formation region (e.g. the Perseus molecular cloud), and $2.5$\,kpc to mimic a distant star formation region such as those seen in the GLIMPSE survey (e.g. the Eagle Nebula).
We used the fluxes integrated in the aperture closest to a 3'' radius aperture at these distances (i.e. $770$\,AU and $7130$\,AU).
We then applied sensitivity limits, using the bright and faint limits listed in Table~\ref{t:glimpse_sens}.
The JHK limits are typical 2MASS values.
The IRAC and \mips limits are similar to those quoted for several large-scale surveys, such as the c2d \citep{harvey06,jorgensen06} and SAGE \citep{meixner06} surveys.

These limits will vary with exposure time and background levels, and the limits given here are just an example of a plausible range.
The distribution of masses and ages remaining after applying these limits to the IRAC bands are shown in Figure~\ref{f:realcluster}: observations of the distant star-forming cluster with these sensitivity and saturation limits would be less sensitive to low-mass YSOs and more sensitive to high-mass YSOs than observations of the nearby cluster.
In addition, observations of the distant star-forming cluster would be slightly biased towards earlier stages of evolution, as the luminosity of pre-main sequence stars decreases with age.

\subsubsection{Color-color plots}

Color-color plots have been widely used to classify YSOs, including for example JHK and IRAC color-color plots.
In a recent study of this color-color space, \citet{allen04} proposed that disk-only sources should fall mostly in a box defined by  $0.4<([5.8]-[8.0])<1.1$ and $0.0<([3.6]-[4.5])<0.8$, whereas younger sources with infalling envelopes should fall redwards of this location.
Although the grid of models that was used covered a range of temperatures and evolutionary states, the interpretation of the color-color diagram was made using only models with a temperature of $4,000$\,K, an age of $1$\,Myr, and using only one inclination.
As shown by \citet{hartmann05}, this analysis is appropriate for the Taurus star formation region where most sources have stellar temperatures ranging from $3,000$\,K to $5,000$\,K
\citep{kh95}.
However, as mentioned by \citeauthor{hartmann05}, some of the very young sources, such as IRAS 04368+2557, have bluer colors than predicted by \citeauthor{allen04}.
Further study of this color-color space is needed to investigate whether such a classification can apply to more distant and massive star formation regions where the sources seen are likely to have a much wider range of ages and temperatures. 
In this section we re-examine IRAC color-color space for nearby and distant clusters, and present results for other color-color spaces.

In Figure~\ref{f:o1} we show the distribution of the entire grid of radiation transfer models in JHK (J-H vs H-K), IRAC ([3.6]-[4.5] vs [5.8]-[8.0]), and IRAC+\mips ([3.6]-[5.8] vs [8.0]-[24.0]) color-color plots.
The plots show the number of models in a logarithmic grayscale for all models in the grid, as well as for each individual Stage (I/II/III).
Figure~\ref{f:op} shows the fraction of models at each Stage relative to the total, for the same color-color spaces as Figure~\ref{f:o1}.
Dark areas show where most of the models correspond to a given Stage: for example a dark area in a `Stage~I/All' ratio plot indicates a region where most models are Stage~I models.
To ensure that these results are not biased by unrealistic models, we show the same plots for our simulated clusters with sensitivity and saturation limits applied, at $250$\,pc  (Figures \ref{f:c11} and \ref{f:c1p}) and $2.5$\,kpc (Figures \ref{f:c21} and \ref{f:c2p}).
The main difference between color-color plots for the entire grid and for the virtual clusters is that there are many fewer Stage~I models in the virtual cluster plots than in the entire grid.
This is expected as the ages are distributed logarithmically in the entire grid, and linearly in the cluster.
In addition, faint Stage~I models will be removed due to the applied sensitivity limits.

\paragraph{JHK color-color plots} 
The models in our grid tend to lie along and redward in (H-K) of the locus for reddened stellar photospheres.
For the whole grid and for the two virtual clusters, a number of Stage~I models occupy a small region where no Stage~II and III models lie.
However the region where only Stage~I models are found is different in each case.
Stage~II models also lie along and redward in (H-K) of the locus for stellar photospheres, while the colors of Stage~III models are in most cases identical to stellar photospheres.

Observations of distant star formation regions are likely to be affected by high levels of extinction that would make most YSOs along the locus of reddened stellar photospheres indistinguishable from highly reddened stars.
This suggests that there are in fact no regions in (J-H) vs. (H-K) color-color space where only sources at a specific stage of evolution always lie, and therefore that (J-H) vs. (H-K) colors are not a reliable indicator of the evolutionary stage of a source.

YSOs displaying colors with a redder (H-K) color than reddened stellar photospheres can still be classified as such, but their evolutionary stage cannot be determined, while many YSOs will be simply be indistinguishable from reddened stellar photospheres.

\paragraph{IRAC color-color plots}
The models in our grid lie mostly redward in [3.6]-[4.5] and [5.8]-[8.0] compared to stellar photospheres, which fall mostly at (0,0).
The Stage~I models in our grid occupy a large region of IRAC color-color space, which includes a substantial region unoccupied by Stage~II and III models.
Many Stage~I models are fairly red at [3.6]-[4.5], but are not always as red at [5.8]-[8.0] as the predictions by \citet{allen04}.
The main difference in our models is that we include bipolar cavities that allow more scattered light to emerge, which tends to make the sources bluer in this color.
Furthermore, pole-on Stage~I models tend to have bluer colors, as one can view the star unobscured by the envelope, by looking down the cavity.
The region where only Stage~I models fall is similar for the whole grid of models and for the virtual clusters.
Most of our Stage~II models lie in the same region indicated by \citeauthor{allen04} as the ``disk domain''.
However, many also lie outside this region, due to variations in disk mass, stellar temperature, and inner hole size, as discussed below.
Most regions occupied by Stage~II and III models are also occupied by Stage~I models.
This suggests that although many Stage~I sources can be identified uniquely as so, the evolutionary stage of the remaining Stage~I sources as well as of most Stage~II and Stage~III sources cannot be reliably found from the IRAC colors alone.
This can be seen in Figures~\ref{f:op}, \ref{f:c1p}, and \ref{f:c2p}, which shows that the `Stage~I/All' fraction is close to $100$\,\% (black areas) in a large region  of IRAC color-color space, while the `Stage~II/All' and `Stage~III/All' fractions never reach $100$\,\% (grey areas).
We note that most sources that fall in the \citeauthor{allen04} ``disk domain'' are most likely to be Stage~II sources, but can in some cases be Stage~I sources. 

In Figure~\ref{f:schematic} (and overplotted on the IRAC color-color plots in Figures \ref{f:op} through \ref{f:c2p}) we show the approximate regions corresponding to the different evolutionary stages.
These should of course be seen only as general trends.
For example, one complication for Stage~I identification is that the Stage~I region lies along the reddening line from the Stage~II region.
However, as shown by the reddening vector, substantial amounts of extinction ($\av>20$) are required to change the colors of a source significantly.

This suggests that unlike JHK color-color plots, IRAC color-color plots appears to be effective in separating stars with no circumstellar material from most Stage~I and II sources (the colors of many Stage~III sources are very similar to those of stars in the IRAC bands).
Furthermore, very young (Stage~I) sources can be distinguised in many cases from Stage~II/III sources.

\paragraph{IRAC+\mips color-color plots}
As for IRAC color-color plots, the models in our grid lie mostly redward in [3.6]-[5.8] and [8.0]-[24.0] compared to stellar photospheres, which fall mostly at (0,0).
Stage~I models cover a very wide range of colors, and a large number fall in regions that are not occupied by Stage~II and III models.
Furthermore, Stage~II and III models also seem to separate into well-defined regions, suggesting that IRAC+\mips color-color plots may be effective in discriminating between various evolutionary stages.

As before, the approximate regions corresponding to Stage~I, II, and III models are shown in Figure~\ref{f:schematic}, and are overplotted on the IRAC+\mips color-color plots.
We note that even if a source is not detected in \mipsns, an upper limit on its flux can still provide constraints on its evolutionary stage: for example, if a source has [8.0]-[24.0]$<2$, it is not likely to be a Stage~I source.

As in Section~\ref{s:sivsparams}, we find once again that that including data at wavelengths longward of $20$\,\microns is valuable in assessing the evolutionary stage of YSOs.

\subsubsection{Colors and physical parameters}

In Figures \ref{f:cc_mdote} to \ref{f:cc_rmine}, we explore how various physical parameters affect the colors of our models.

\paragraph{Envelope accretion rate} Figure~\ref{f:cc_mdote} shows how the colors of our models depend on $\mdote/\mstar$.
Interestingly, the models with very high envelope accretion rates are relatively blue in JHK, IRAC and IRAC+MIPS color space compared to models with lower accretion rates.
This is due to the complete extinction of the stellar and inner disk/envelope radiation, leaving only scattered light that is relatively blue \citep{whitney03p2,whitney03p1}.
These may however be very faint and thus below detection limits at large distances.
For example the bluest models in [5.8]-[8.0] in the IRAC color-color diagram for the whole grid (Figure~\ref{f:o1}) do not appear in the same diagram for the virtual cluster at $2.5$\,kpc after sensitivity limits have been applied (Figure~\ref{f:c21}).
However, even in this case the youngest models still display colors bluer than would be expected if bipolar cavities had not been included.
In the same way as with spectral indices (c.f. Section~\ref{s:sivsparams}), models with accretion rates $\mdote/\mstar<10^{-6}$\,yr$^{-1}$ have colors similar to disk-only models; i.e. the envelope does not dominate the near- and mid-IR colors.

\paragraph{Disk mass} Figure~\ref{f:cc_mdisk} shows the effect of disk mass on the colors of Stage~II models.
The effect on the JHK, IRAC, and \mips colors is negligible.
This is not surprising, since the disk is opaque at these wavelengths, and the near- and mid-IR radiation originates only from the surface of the disk (as previously found in Section \ref{s:sivsparams}).

\paragraph{Stellar Temperature} Figures \ref{f:cc_tstar1} and \ref{f:cc_tstar2} show the effect of stellar temperature on the colors of Stage~I
and Stage~II models respectively.
\citet{whitney04} showed that the IRAC colors of YSOs become redder for hotter stellar sources.
These figures show that this applies more generally to near- and mid-IR colors.
This echoes the result found in Section \ref{s:sivstemp} that the spectral index of our models calculated using near-IR and mid-IR wavelengths increases with stellar temperature.

\paragraph{Disk and envelope inner radius} Figure~\ref{f:cc_rmine} shows the effect of increasing the inner radius of the disk and envelope on the colors of all the models in the grid.
Since the inner holes are completely evacuated, this means that as the inner radius of the disk is increased, the temperature of the warmest dust decreases, progressively removing flux from shorter to longer wavelengths; thus only photospheric fluxes remain at shorter wavelengths, where the disk contribution has been removed.

The colors of the models in the JHK color-color plot tend to the photospheric colors for $10$\,$\rsub < \rmind < 100$\,$\rsub$ (note that photospheric colors are not necessarily (J-H)=0 and (H-K)=0 except for high photospheric temperatures).
For these values of the inner radius, the [3.6]-[4.5] colors and the [3.6]-[5.8] colors also tend to the photospheric colors (close to or equal to zero), while the [8.0]-[24.0] colors are not significantly affected.
For inner radii $\rmind>100$\,$\rsub$, most models have [3.6]-[4.5] and [3.6]-[5.8] equal to zero, and the [5.8]-[8.0] colors also tend to zero (as can be seen from the high concentration of models at (0,0) in the IRAC color-color plot).
The models with $[3.6]-[4.5] > 0$ or $[3.6]-[5.8] > 0$ and with $\rmind>100$\,$\rsub$ are typically embedded sources for which the extinction to the central source produces colors that are redder than stellar photospheres.
We note that the only regions of color-color space where models with holes can be unambiguously identified as such are the regions for which [3.6]-[4.5]=0 (e.g. in the IRAC color-color plot) and [3.6]-[5.8]=0 (e.g. in the IRAC+MIPS color-color plots).
These regions are shown in Figure~\ref{f:schematic}.

\section{Discussion and Conclusion}
\label{s:conclusion}

We have computed a large grid of SEDs from axisymmetric YSO models using a Monte-Carlo radiation transfer code.
These models span a large range of evolutionary stages, from the deeply embedded protostars to stars surrounded only by optically thin disks.

We have made the 20,000 models publicly available on a dedicated web server\footnote{http://www.astro.wisc.edu/protostars}.
For each model, the following output are available for download:
\begin{itemize}
\item The SEDs for 10 inclinations (from pole-on to edge-on) and integrated in 50 different circular apertures (with radii between $100$ to $100,000$\,AU).
\item Three separate SEDs, constructed from the energy packets whose last point of origin is the star, the disk, and the envelope respectively (see Section~\ref{s:seds}), for each inclination and aperture. In addition, an SED constructed for each inclination and aperture from photons who last scattered before escaping, and an SED constructed for each inclination from photons which escaped directly from the stellar surface are available.
\item A polarization spectrum for each inclination and aperture.
\item Convolved fluxes and magnitudes for each inclination and aperture for a wide range of filters.
\end{itemize}
Users have the option of downloading the convolved fluxes and the model parameters for all models as single files.
These files can be used for example to carry out the analysis of color-color spaces or spectral indices not covered in this paper, or to compare the models to data directly.

We request feedback from the community, observers and theorists alike, on how to improve our grid of models, as we plan to run new grids of models over the next few years.
For example, we plan to produce higher signal to noise SEDs in the next grid of models, as well as produce resolved images, in addition to the planned model improvements described in Section~\ref{s:caveats}.

We have presented typical model SEDs from our grid (Figure~\ref{f:seds}), including separate SEDs constructed from the energy packets whose last point of origin are the star, the disk and the envelope (Figure~\ref{f:seds_indiv}).
We have also presented polarization spectra for these models in UBVRI and JHKL bands (Figure~\ref{f:seds_pol}). 
In addition, we have shown the variation of the K-band polarization for all of our models with disk mass and envelope accretion rate (Figure~\ref{f:k_pol}).
Stage~I sources show a high polarization ($>5$\,\%) for envelope accretion rates above $\mdote/\mstar\sim2\times10^{-6}$\,yr$^{-1}$, over a large range of viewing angles ($45-90^{\circ}$) and a broad wavelength range ($0.5-10$\,\micronsns).
Stage~I sources can show a $90^{\circ}$  position angle rotation with wavelength, as discussed by previous authors \citep{bastien87,kenyon93p2,whitney97} due to the dominance of scattering in either the cavity or the disk.
Stage~II sources show high polarization ($>3$\,\%) only when viewed edge-on, but show significant polarization ($>0.5$\,\%) over a range of viewing angles.

We have carried out an analysis of JHK, IRAC and MIPS spectral indices and color-color plots using our model SEDs, and have constructed `virtual clusters' in order to understand how a cluster of young stars might look in these color-color spaces.
Our results indicate that:
\begin{itemize}
\item How well a given spectral index indicates the evolutionary stage of a source is dependent on the range of wavelengths of the fluxes used to calculate the spectral index (Figure~\ref{f:sivsparams}).
Our results suggest that the use of fluxes beyond $\sim20$\microns (e.g. \mipsns), in addition to $1-10$\microns fluxes, in calculations of the spectral index of YSOs is valuable for deriving information about the evolutionary stage of these sources.
\item For optically thin disks, the traditional $2.0\rightarrow25.0$ spectral index increases with disk mass, while for optically thick disks, the same spectral index is insensitive to disk mass, as the emergent flux at these wavelengths originates only in the outer layers of the disk (Figure~\ref{f:sivsparams}).
\item The near- and mid-IR colors of a disk-only source are sensitive to stellar temperature, disk inner radius, and disk flaring power (Figure~\ref{f:siadditional}).
In particular, we find that for temperatures above $5,000$\,K, these colors become redder as the temperature increases, while the presence of an inner hole decreases the flux at near- and mid-IR wavelengths.
For large enough holes, the near- and mid-IR colors are the same as those of a stellar photosphere.
The effect of a larger disk flaring power is to increase the amount of flux intercepted by the disk, and thus to increase the amount of reprocessed flux at mid- and far-IR wavelengths. 
\item For a given source, the spectral index is sensitive to the wavelength range of the broadband fluxes used to calculate the spectral index (Figure~\ref{f:siranges}), and to a lesser extent is also dependent on whether the spectral index is calculated from the slope of a line joining two points, or as the slope of a least-squares fit line to a larger number of fluxes (Figure~\ref{f:sidensity}).
\item (J-H) vs. (H-K) color-color plots can be used to identify sources as YSOs if they lie redward in (H-K) of the locus for reddened stellar photospheres, although their evolutionary stage cannot be reliably found simply from the JHK colors.
In addition, many YSOs are likely to be indistinguishable from reddened stellar photospheres in regions of high extinction.
\item Stage~I models tend to occupy large regions in IRAC and IRAC+\mips color-color plots, due to inclination effects, scattering in the bipolar cavities (which makes some edge-on sources blue), and variations in stellar temperature.
\item $[3.6]-[4.5]$ vs. $[5.8]-[8.0]$ IRAC color-color plots contain a large region that is likely to be occupied only by Stage~I sources.
The region corresponding to \citet{allen04}'s ``disk domain'' is likely to contain mostly Stage~II sources, but also a number of Stage~I sources.
The remaining regions in this color-color space can be occupied by sources at {\it any} evolutionary stage.
\item $[3.6]-[5.8]$ vs. $[8.0]-[24.0]$ IRAC+\mips color-color plots provide a good separation of YSOs at different evolutionary stages.
We note that even if a given source is not visible in \mipsns, an upper limit on the flux can still constrain its evolutionary stage.
\end{itemize}

In Figure~\ref{f:schematic} we show approximate regions in IRAC and IRAC+\mips color-color space where models of a given evolutionary stage lie, irrespective of whether we are using the whole grid of models or the virtual clusters.
In summary we find that near-IR fluxes (such as JHK fluxes) can be used to discriminate between reddened stars and many (but not necessarily all) YSOs, but cannot reliably determine the evolutionary stage of a YSO.
Mid-IR fluxes (such as IRAC fluxes) should be efficient in separating YSOs from stellar photospheres, and allow some of the youngest sources to be unambiguously classified as such, although the evolutionary stage of the remaining YSOs is likely to be ambiguous.
Finally, including fluxes at wavelengths beyond $\sim20$\microns (such as \mipsns) in addition to near- and mid-IR fluxes substantially improves the ability to distinguish between the various evolutionary stages of YSOs.

\acknowledgments

We wish to thank Alexander Lazarian, Mike Wolff, and the GLIMPSE team for allowing us to use their respective computer clusters.
We benefited from discussions with Ed Churchwell and Ian Bonnell.
The referee, Klaus Pontoppidan, provided useful suggestions that improved the paper.
Support for this work was provided by NASA's Long Term Space Astrophysics Research Program (NAG5-8412) (BW, TR), the Spitzer Space Telescope Legacy Science Program through contract 1224988 (BW, TR), NASA's Spitzer Space Telescope Fellowship Program (RI), a Scottish Universities Physics Alliance Studentship (TR), and a PPARC advanced fellowship (KW), and the NSF REU Site grant 0139563 to the University of Wisconsin - Madison (PD).


\bibliography{}


\clearpage

\begin{deluxetable}{cl}
\tablewidth{0pt}
\tablecaption{The 14 parameters varied for the 20,000 YSO models\label{t:parameters}}
\tablehead{\colhead{Symbol} & \colhead{Description}}
\startdata
$\mstar$    &  Stellar Mass                \\
$\rstar$    &  Stellar Radius              \\
$\tstar$    &  Stellar Temperature         \\
$\mdote$    &  Envelope accretion rate     \\
$\rmaxe$    &  Envelope outer radius       \\
$\rhoconst$ &  Cavity density              \\
$\thetacav$ &  Cavity opening angle        \\
$\mdisk$    &  Disk mass (gas+dust)        \\
$\rmaxd$    &  Disk outer radius           \\
$\rmind$    &  Disk inner radius           \\
$\mdotdisk$ &  Disk accretion rate         \\
$\zmin$     &  Disk scaleheight factor     \\
$\beta$     &  Disk flaring angle          \\
$\rhoamb$   &  Ambient density
\enddata
\end{deluxetable}

\clearpage

\begin{deluxetable}{ccc}
\tablewidth{0pt}
\tablecaption{The bright and faint limits used to create the virtual cluster\label{t:glimpse_sens}}
\tablehead{\colhead{Band} & \colhead{Bright limit} & \colhead{Faint limit}}
\startdata
J & 8.0 & 16.5 \\
H & 7.0 & 15.5 \\
K & 6.5 & 15.0 \\
$[3.6]$ & 6.0 & 18.0 \\
$[4.5]$ & 5.5 & 17.0 \\
$[5.8]$ & 3.0 & 15.0 \\
$[8.0]$ & 3.0 & 14.0 \\
$[24.0]$ & 0.6 & 10.0
\enddata
\end{deluxetable}


\clearpage

\clearpage

\clearpage

\clearpage

\clearpage

\begin{figure}
\epsscale{0.38}
\plotone{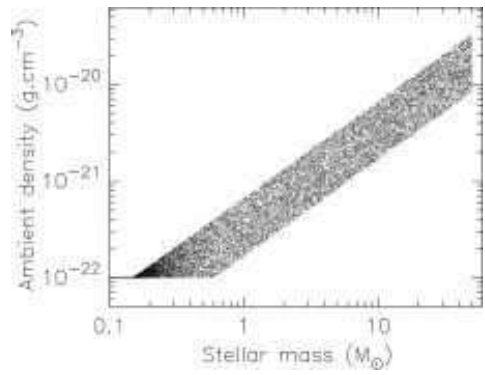}
\caption{The ambient density as a function of the central source mass $\mstar$. A lower limit of $10^{-22}$\,g\,cm$^{-3}$ was used.}
\end{figure}

\clearpage

\begin{figure}
\epsscale{0.7}
\plotone{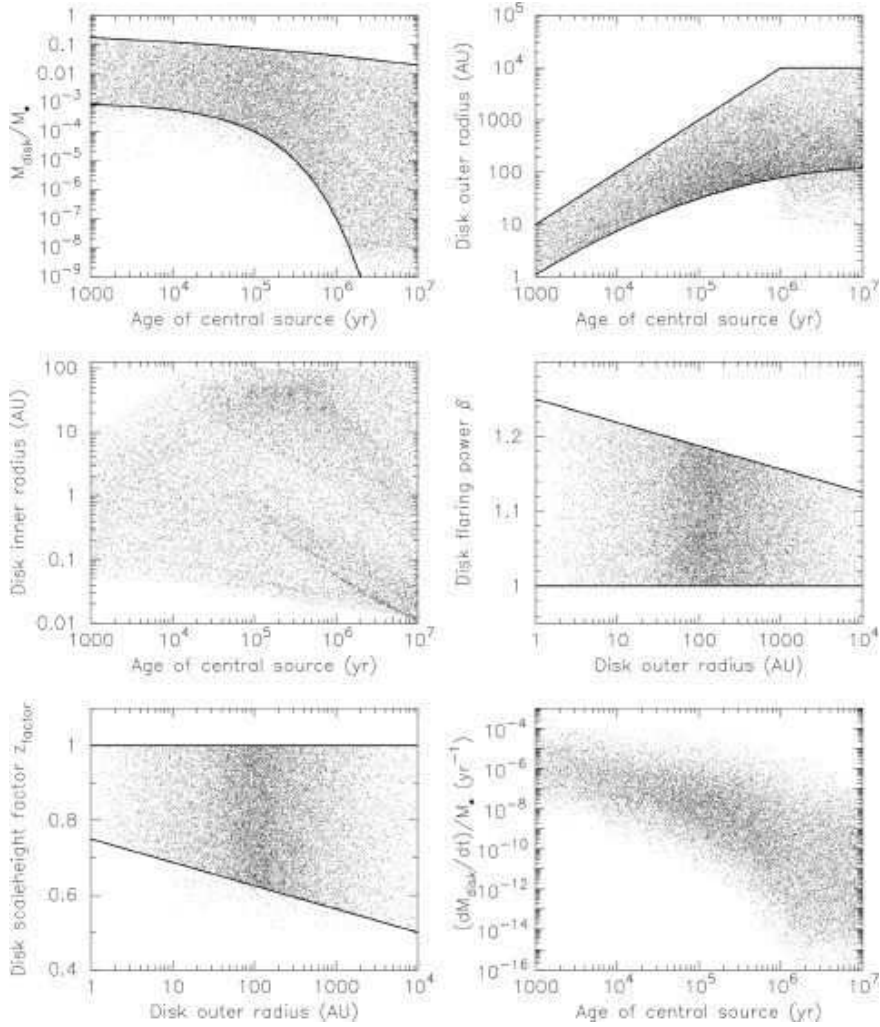}
\caption{The disk parameters as a function of evolutionary age or disk outer radius for the 20,000 models. The disk masses (top left) were initially sampled between the two solid lines, and the points that lie below the lower limit are models for which an inner hole has been cleared in the disk. The disk outer radius (top right) was initially sampled between the two solid lines, and the points that lie below the lower limit for $\agestar>10^6$\,yr are models for which the outer radius was truncated (this was done only for disk-only models). The disk inner radius (center left) was sampled between $\rmind=1\rsub$ and $\rmind=100$\,AU for two thirds of models, and set to $\rmind=1\rsub$ for the remaining models. The flaring power $\beta$ (center right) was sampled as a function of $\rmaxd$ between the two solid lines. The disk scaleheight factor $\zmin$ (bottom left) was sampled as a function of $\rmaxd$ before outer disk truncation from between the two solid lines, and the points below the line are models for which the outer disk radius was truncated, shifting these models to lower values of $\rmaxd$. The disk accretion rate (bottom right) was not sampled directly, as instead we randomly sampled the values of $\alpha_{\rm disk}$ between $10^{-3}$ and $10^{-1}$. This plot shows the values of $\mdotdisk/\mstar$ calculated by the code (using equation \ref{e:mdotdisk}).\label{f:disk}}
\end{figure}

\clearpage

\begin{figure}
\epsscale{1.0}
\plotone{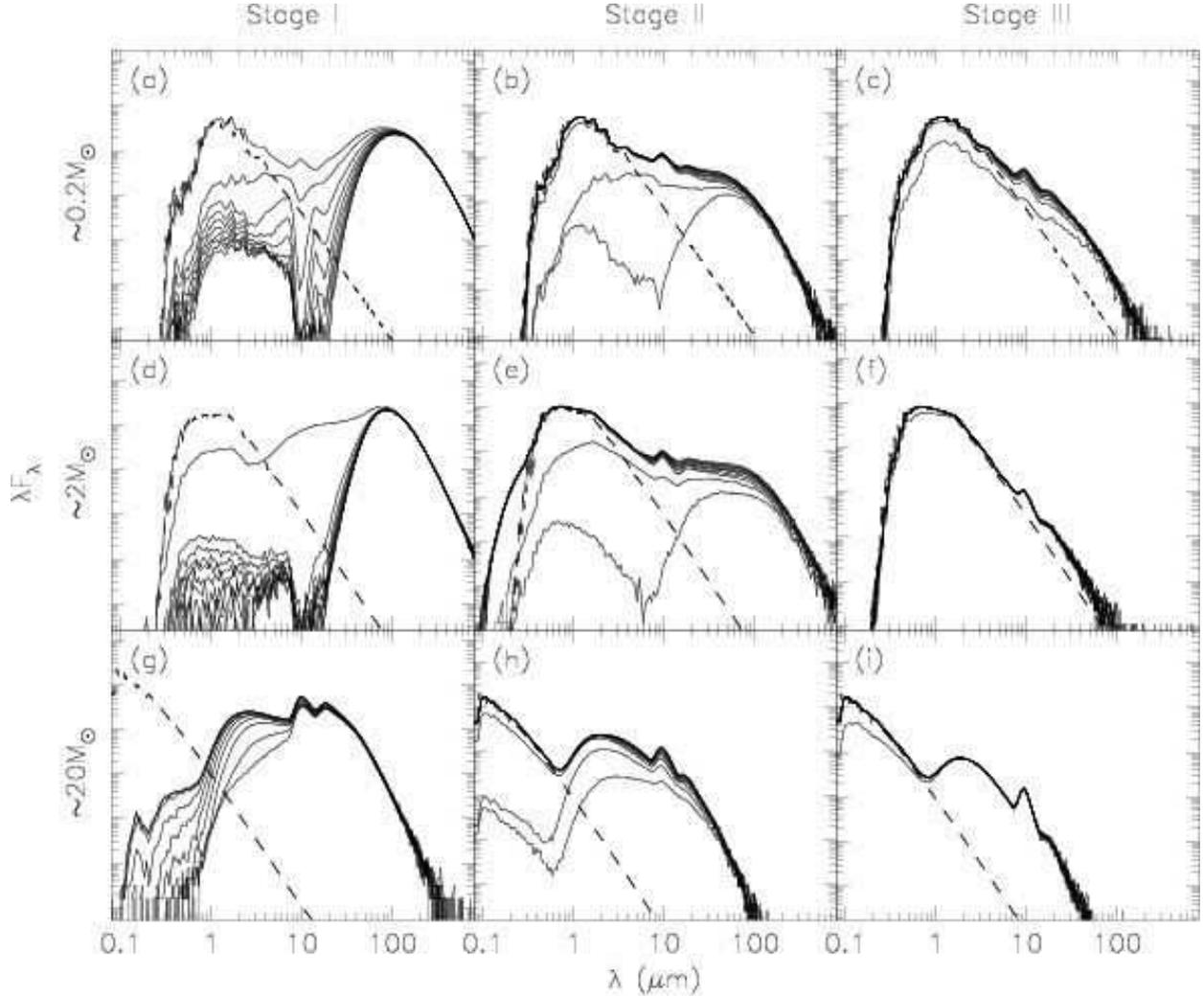}
\caption{Example SEDs from the grid.
Plots (a), (b) and (c) show SEDs for three different $\sim0.2\msun$ objects at Stages I, II and III of evolution respectively.
Plots (d), (e) and (f) show SEDs for three different $\sim2\msun$ objects at Stages I, II and III of evolution respectively.
Finally plots (g), (h) and (i) show seds for three different $\sim20\msun$ objects at Stages I, II and III of evolution respectively.
Each plot shows 10 SEDs, one for each inclination computed.
As the stellar masses and evolutionary ages of the models are randomly sampled, the top, center, and bottom panels only show examples of {\it possible} evolutionary sequences.\label{f:seds}}
\end{figure}

\clearpage

\begin{figure}
\epsscale{1.0}
\plotone{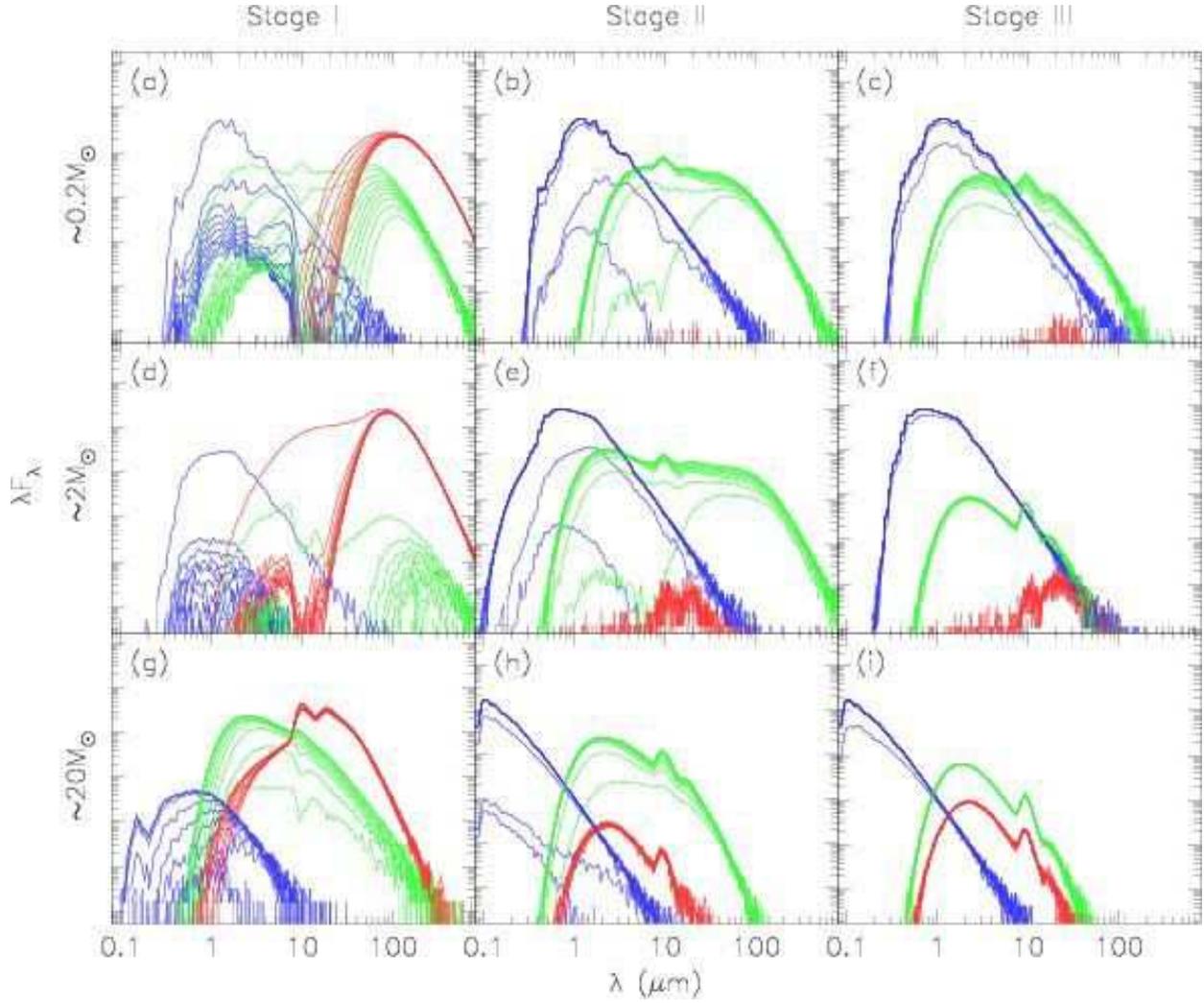}
\caption{The same SEDs as in Figure~\ref{f:seds}, showing the contribution to the SEDs from the energy packets whose last point of origin is the star (blue), the disk (green), and the envelope (red).\label{f:seds_indiv}}
\end{figure}

\clearpage

\begin{figure}
\epsscale{1.0}
\plotone{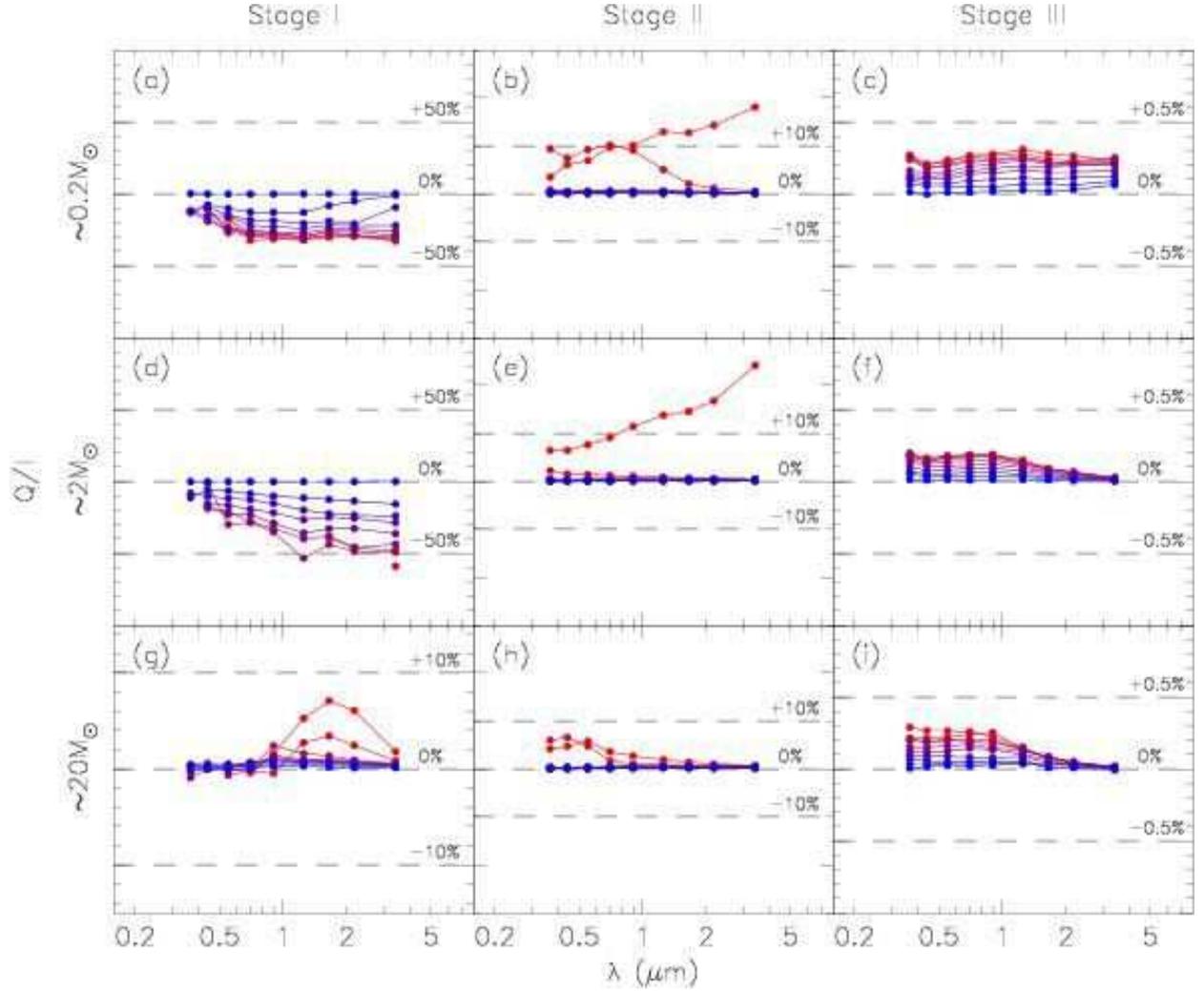}
\caption{The degree of polarization (Q/I) for the same models as in Figure~\ref{f:seds}. Fluxes convolved with the UBVRI and JHKL passband filters are shown. Blue to red colors indicate pole-on to edge-on viewing angles.
Values were not always available for all inclinations due to low signal-to-noise.
Note that the y-axis scale changes from left to right.\label{f:seds_pol}}
\end{figure}

\clearpage

\begin{figure}
\epsscale{1.0}
\begin{center}
\includegraphics[width=4.5in]{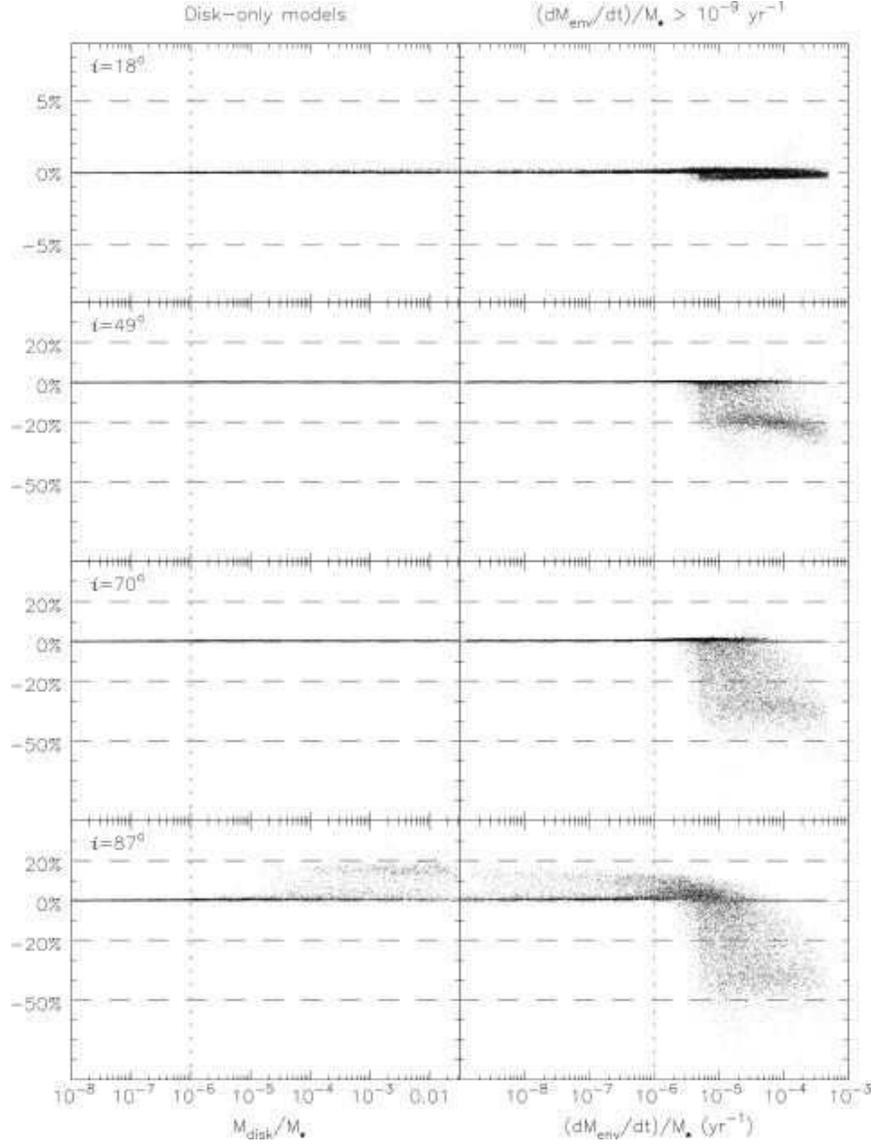}
\end{center}
\caption{The K-band degree of polarization (Q/I) for all models in the grid, shown for four different inclinations, versus the disk mass $\mdisk/\mstar$ for disk-only models (left) and the envelope accretion rate $\mdote/\mstar$ for the remaining models (right).
The vertical dotted lines show the separation between Stage~I (right), II (center) and III (left) models (as defined in Section \ref{s:stage}) on the basis of the physical parameters.
Note that the y-axis scale is expanded in the top panel.
\label{f:k_pol}}
\end{figure}

\clearpage

\begin{figure}
\begin{center}
\includegraphics[width=4.0in]{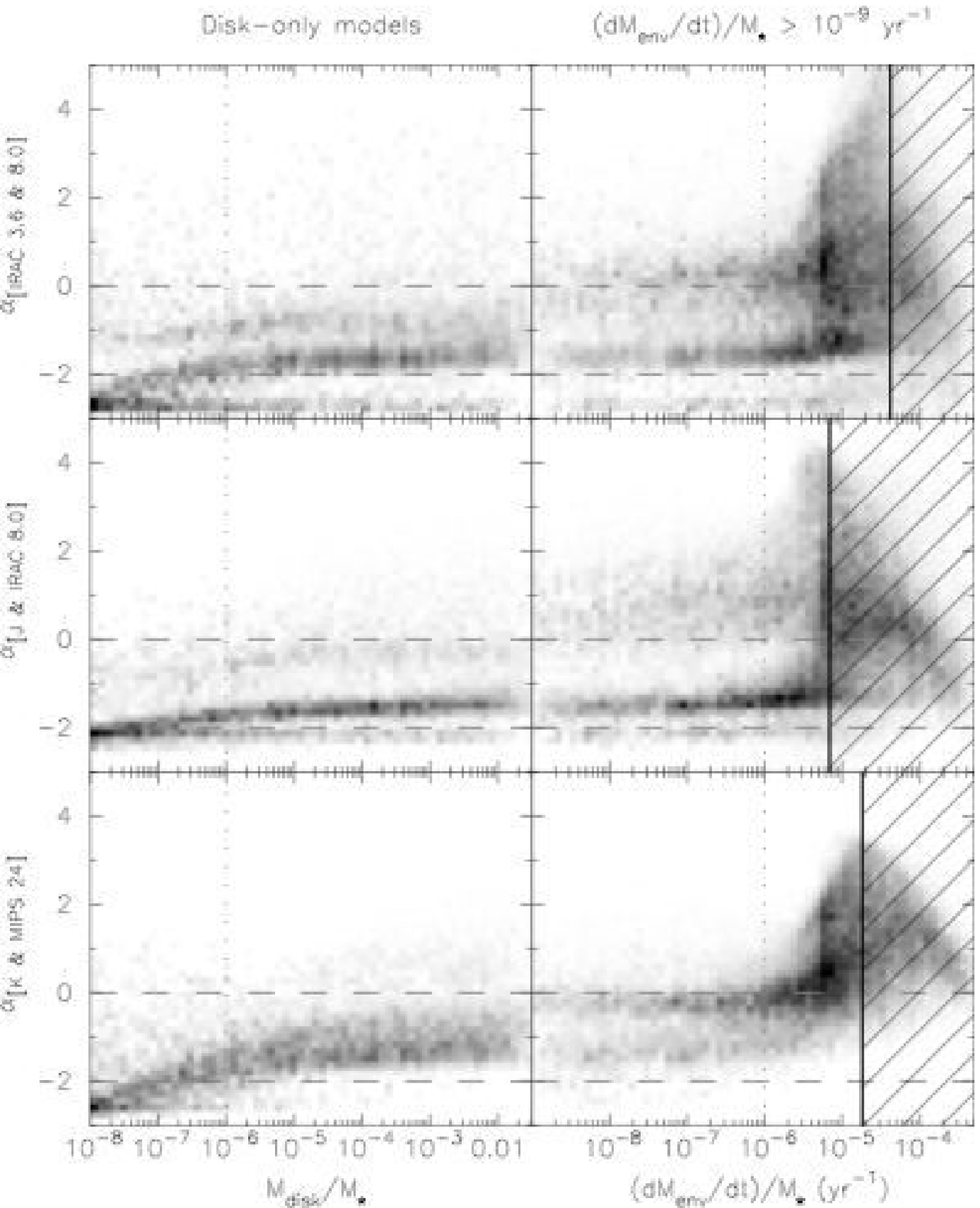}
\end{center}
\caption{Left: spectral indices versus the disk mass $\mdisk/\mstar$ for all disk-only models.
Right: spectral indices versus the envelope accretion rate $\mdote/\mstar$ for all models with $\mdote/\mstar>10^{-9}$\,yr$^{-1}$.
The spectral indices were calculated using IRAC 3.6 \& 8.0\microns fluxes (top), J \& IRAC 8.0\microns fluxes (center), and K \& \mips fluxes (bottom).
The grayscale shows the number of models on a linear scale.
The dashed horizontal lines show the separation between Class~I, II, and III models on the basis of the spectral indices.
The vertical dotted lines are as in Figure~\ref{f:k_pol}: the left and right panel together show the evolution of the spectral index from Stage~III to Stage~I models.
The hashed region shows the values of $\mdisk/\mstar$ for which at least $10$\,\% of SEDs have insufficient signal-to-noise in order to compute a spectral index; the trend of decreasing spectral index inside these regions is due to signal-to-noise limitations.
\label{f:sivsparams}}
\end{figure}

\clearpage

\begin{figure}
\begin{center}
\includegraphics[width=5in,angle=270]{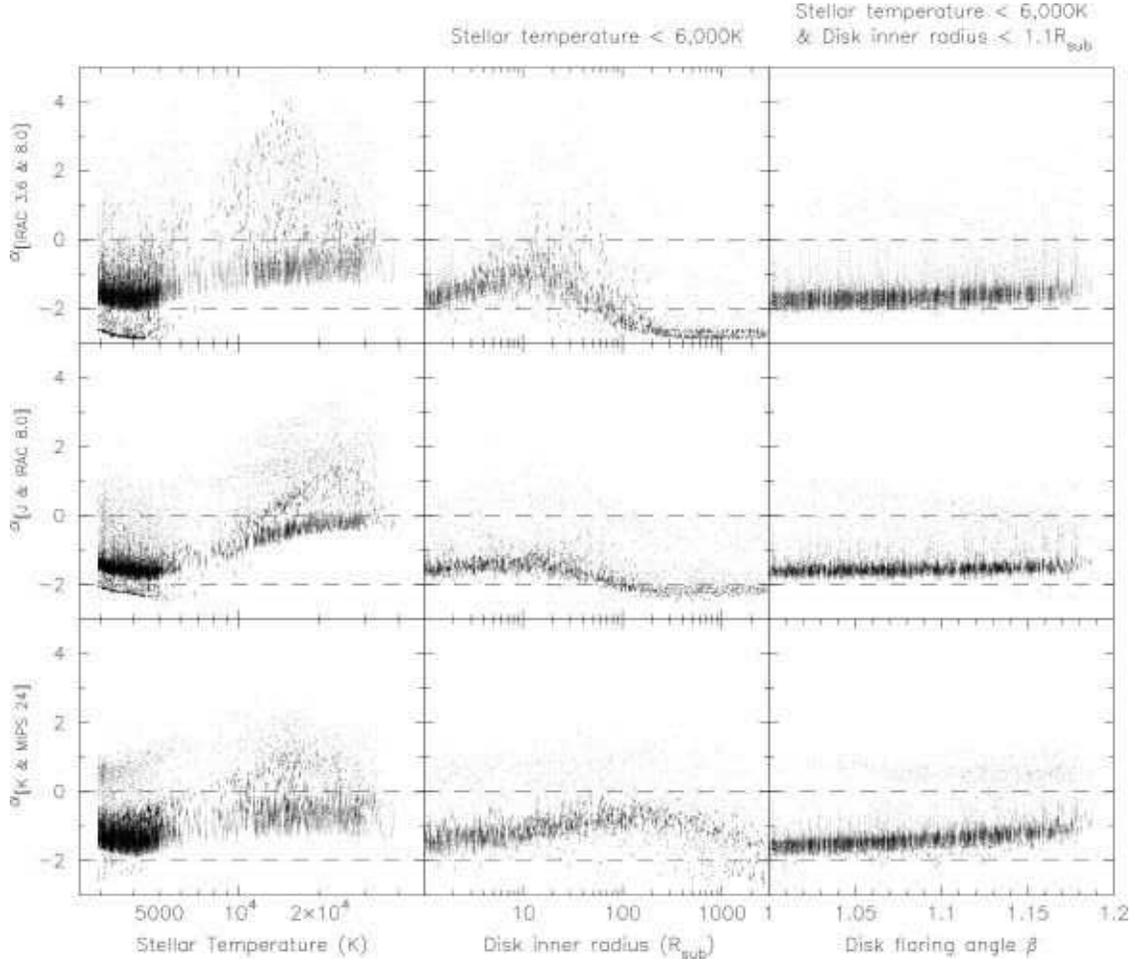}
\end{center}
\caption{Left: spectral indices versus the stellar temperature $\tstar$ for all disk-only models with $\mdisk/\mstar>10^{-5}$.
Center: spectral indices versus the disk inner radius $\rmind$ for all disk only models with $\mdisk/\mstar>10^{-5}$ and with $\tstar<6,000$\,K.
Right: spectral indices versus the disk flaring power for all disk only models with $\mdisk/\mstar>10^{-5}$, with $\tstar<6,000$\,K, and with $\rmind=1\,\rsub$.
The spectral indices were calculated as for Figure~\ref{f:sivsparams}.
The dashed horizontal lines are identical to those shown in Figure~\ref{f:sivsparams}.
\label{f:siadditional}}
\end{figure}

\clearpage

\begin{figure}
\epsscale{1.0}
\plotone{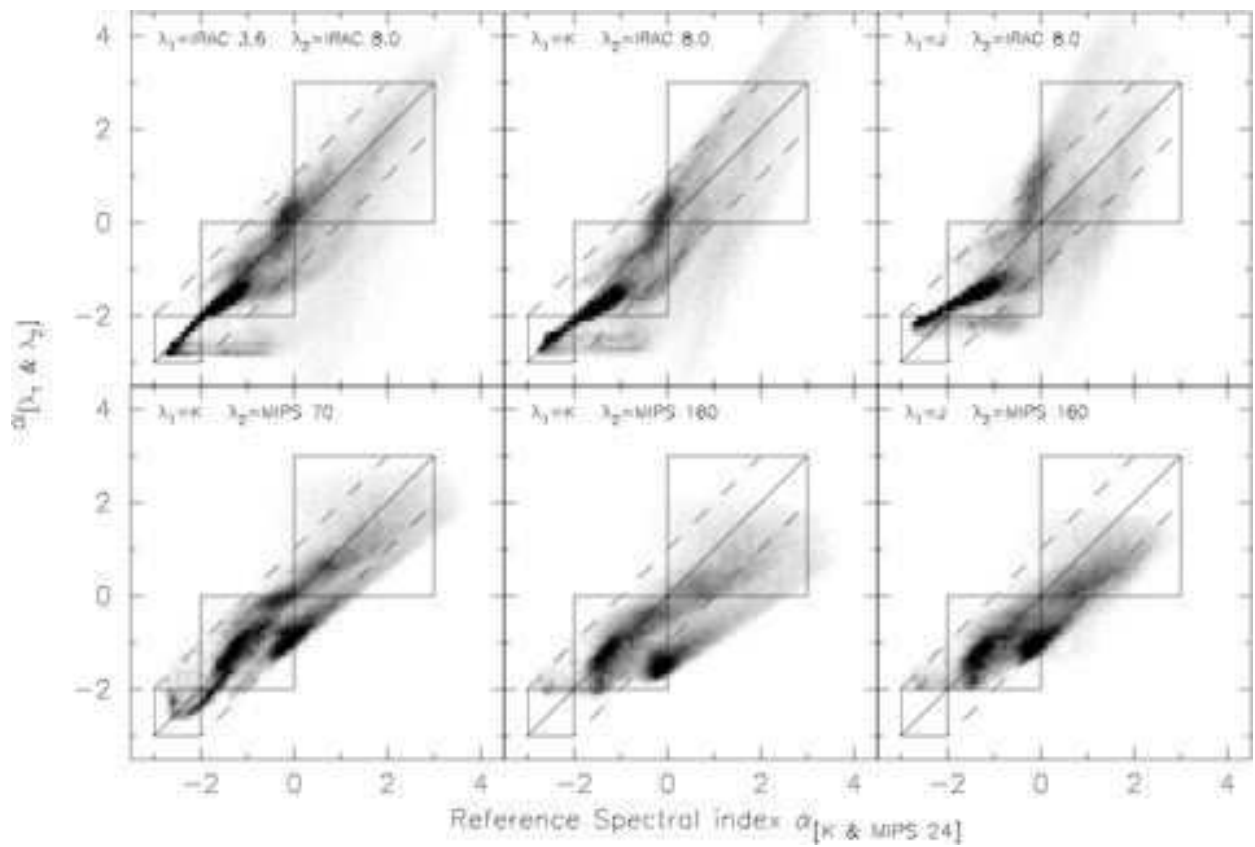}
\caption{The spectral indices for all our models calculated using a \alphaend spectral index (where the values of $\lambda_1$ and $\lambda_2$ are shown in the panels) versus the reference spectral index \alpharefns.
The grayscale shows the number of models on a linear scale.
The solid diagonal line represents the line along which the spectral indices are equal.
The dashed lines encompass the region within which the difference between the spectral indices in less than 1.
The boxes represent regions where the `Class' assigned to a model is the same for the two spectral indices (in each panel, the boxes are from left to right: Class~III, Class~II, and Class~I models).
\label{f:siranges}}
\end{figure}

\clearpage

\begin{figure}
\epsscale{0.7}
\plotone{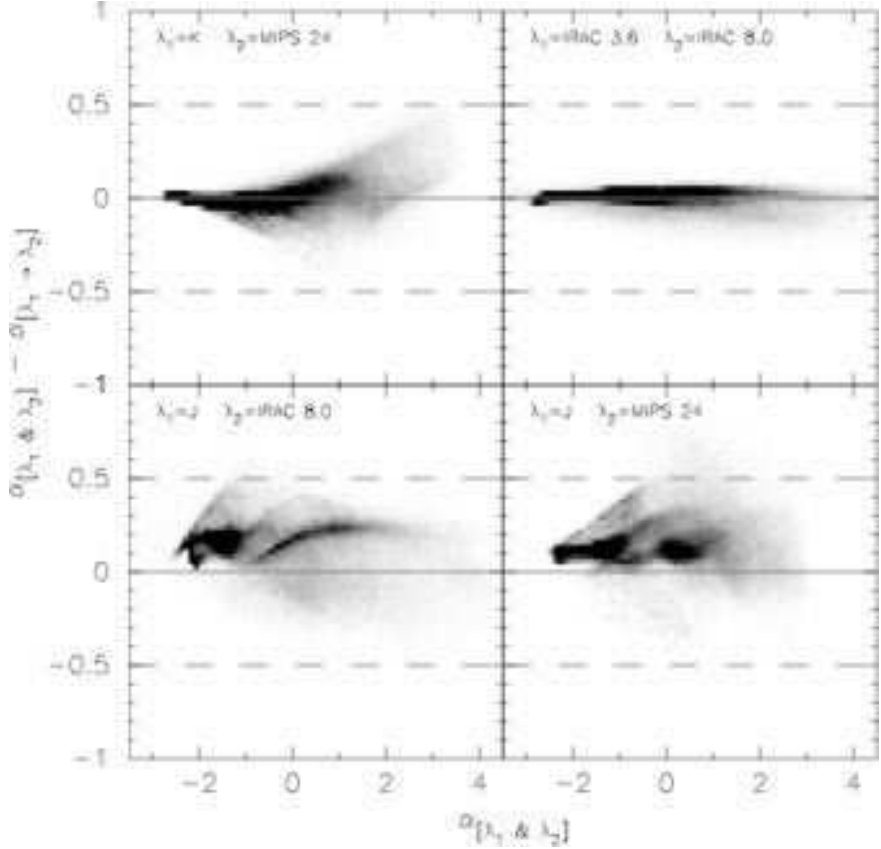}
\caption{The difference between the spectral indices for all our models calculated using only two photometric points (\alphaend) and the spectral indices for the same models calculated by fitting a line to more than two datapoints in the same range (\alphaall), versus the spectral indices calculated using only two photometric points.
The broadband fluxes used to calculate the \alphaall spectral indices are:
K+IRAC+\mips ($\alpha_{[K\,\rightarrow\,\rm{MIPS}\,24]}$),
all IRAC fluxes ($\alpha_{[\rm{IRAC}\,3.6\,\rightarrow\,8.0]}$),
JHK+IRAC fluxes ($\alpha_{[J\,\rightarrow\,\rm{IRAC}\,8.0]}$),
and JHK+IRAC+\mips ($\alpha_{[J\,\rightarrow\,\rm{MIPS}\,24]}$).
The grayscale shows the number of models on a linear scale.
The solid line shows the line along which the spectral indices are equal.
The dashed lines show the line along which the two spectral indices differ by $\pm0.5$.
\label{f:sidensity}}
\end{figure}

\clearpage

\begin{figure}
\epsscale{1.0}
\plottwo{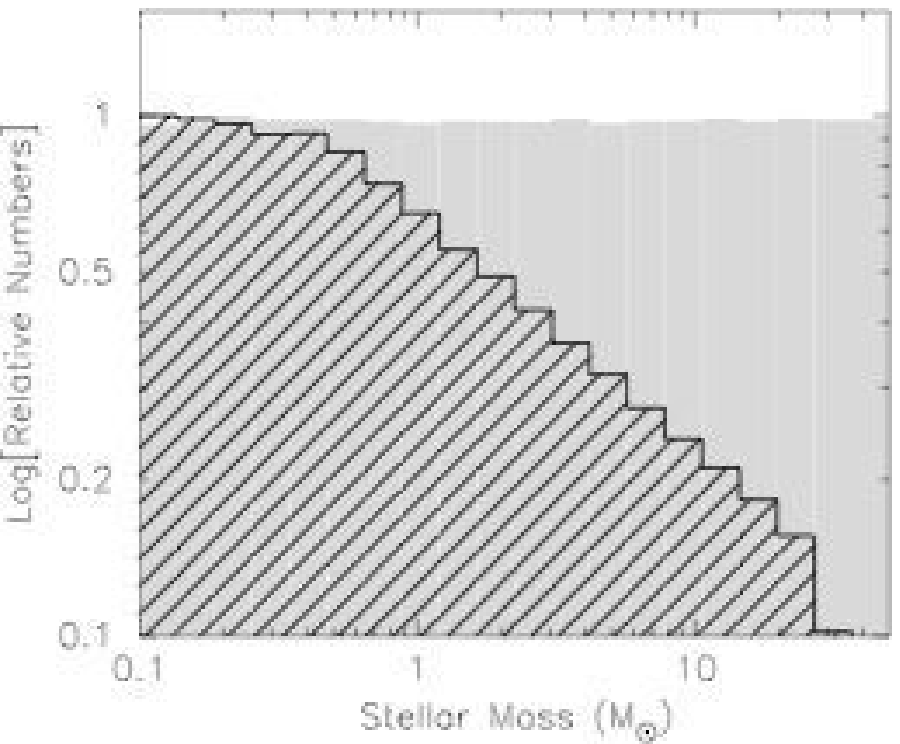}{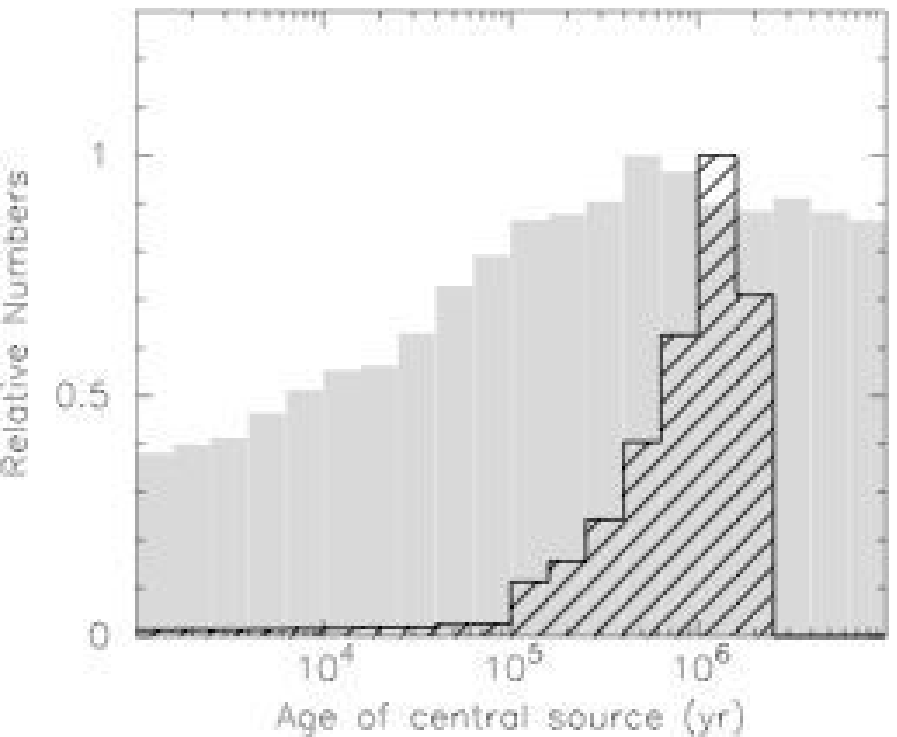}
\caption{The distribution of stellar masses (left) and ages (right) for the entire grid of models (filled gray area) and the virtual cluster before sensitivity cutoffs (hashed area). The stellar masses are sampled logarithmically in the original grid of models, and from a Kroupa IMF in the virtual cluster. The ages are sampled close to logarithmically in the original grid of models, and linearly in the virtual cluster (with a cutoff at $2\times10^6$\,yr).\label{f:cluster}}
\end{figure}

\clearpage

\begin{figure}
\epsscale{1.0}
\plottwo{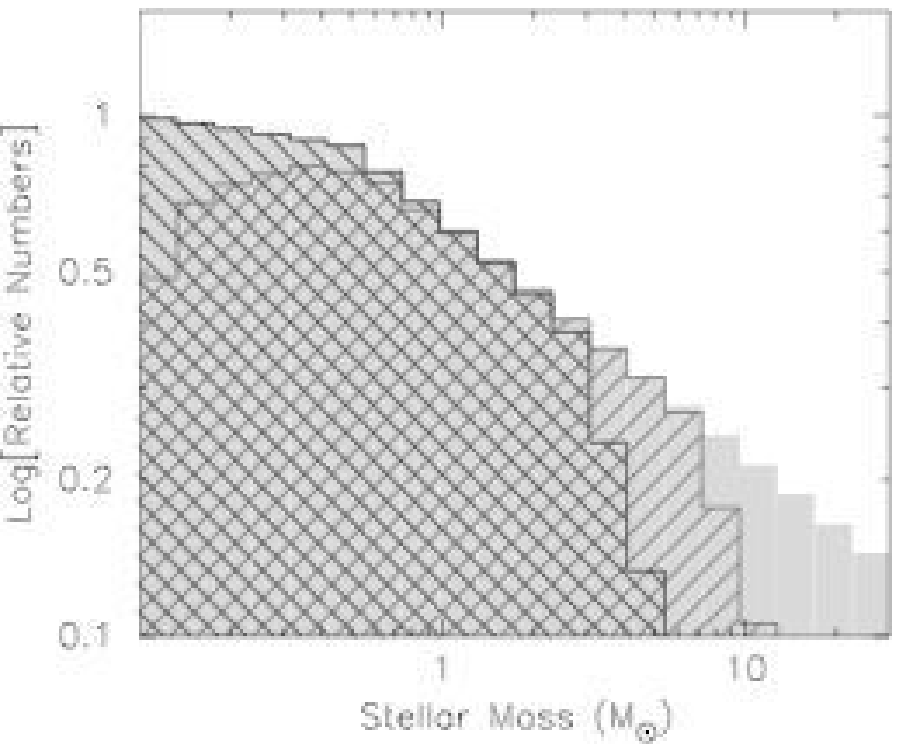}{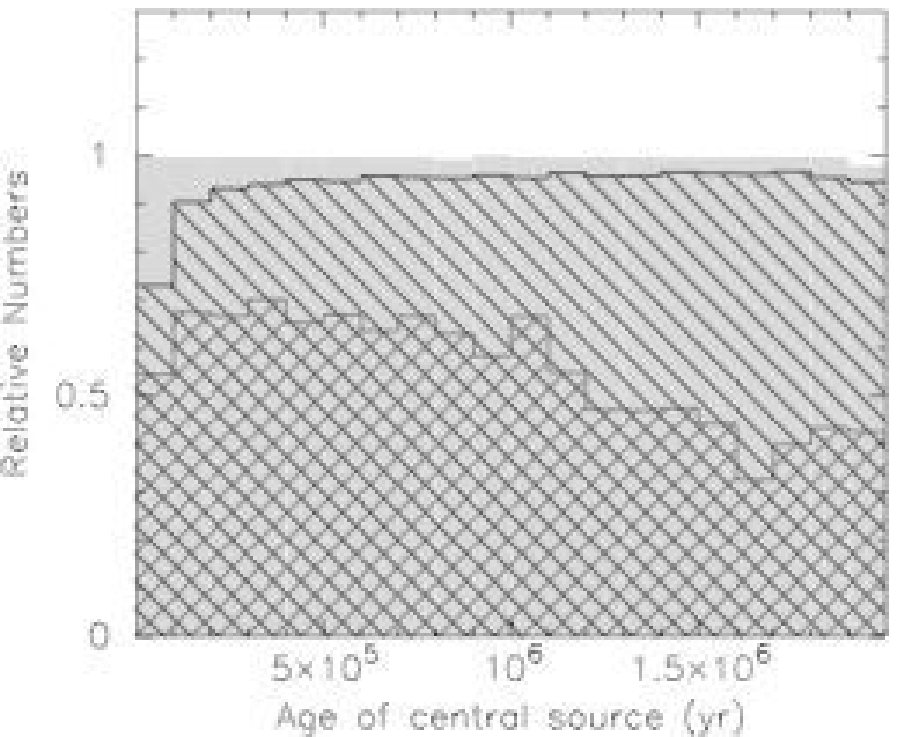}
\caption{The distribution of stellar masses (left) and ages (right) for the virtual cluster before sensitivity cutoffs (filled gray area), after sensitivity cutoffs at $250$\,pc (hashed from top-left to bottom-right) and after sensitivity cutoffs at $2.5$\,kpc (hashed from bottom-left to top-right).
Due to the sensitivity and saturation cutoffs, the distribution of models in the $2.5$\,kpc cluster is more biased towards high mass stars and early evolutionary stages than the $250$\,pc cluster. \label{f:realcluster}}
\end{figure}

\clearpage

\begin{figure}
\epsscale{1.0}
\plotone{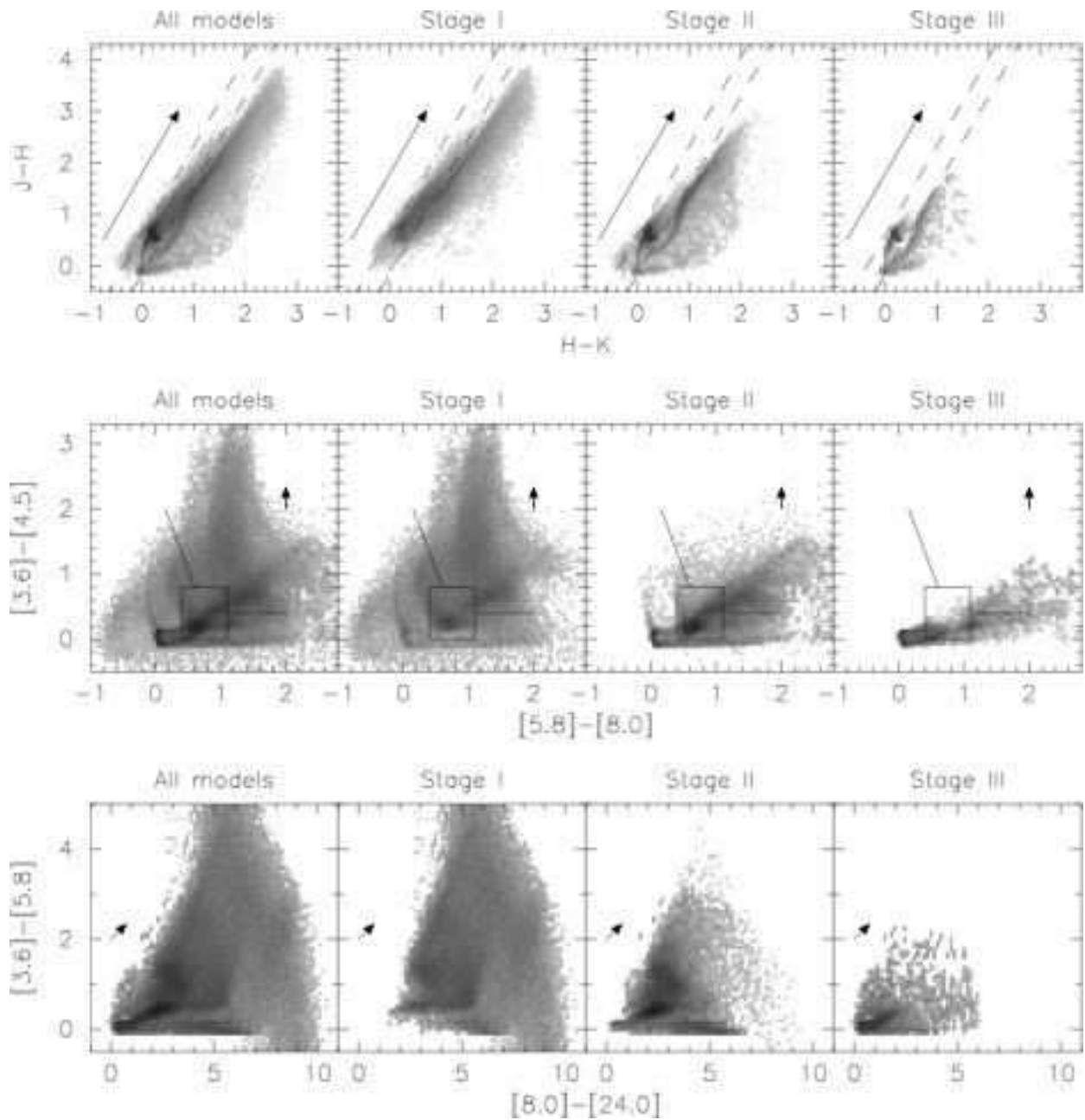}
\caption{JHK, IRAC, and IRAC+\mips color-color plots for all the model SEDs in the grid, showing the number of models on a logarithmic grayscale.
The grayscale is shown down to $0.01$\,\% of the peak value.
The reddening vector shows an extinction of $\av=20$, assuming the \citet{indebetouw05} extinction law. The dashed lines in the JHK color-color plots show the locus of the reddened stellar photospheres. The solid lines in the IRAC color-color plots show the ``disk domain'' and the domain of embedded young objects (redward of the ``disk domain'') from \citet{allen04} and \citet{megeath04}.\label{f:o1}}
\end{figure}

\clearpage

\begin{figure}
\epsscale{0.8}
\plotone{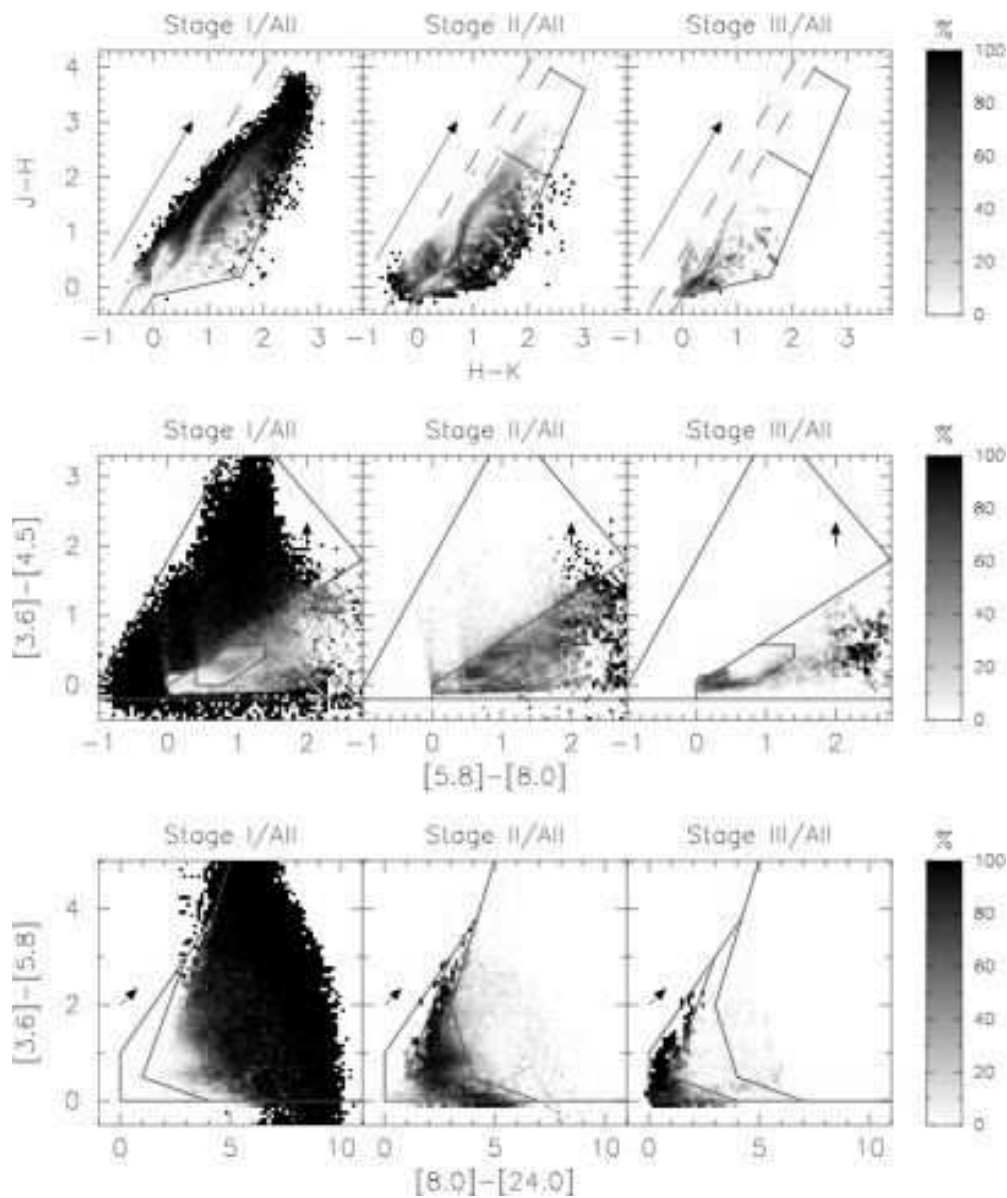}
\caption{Color plots for all the model SEDs showing from left to right: the ratio of the number of Stage~I models to all models, the ratio of the number of Stage~II models to all models, and the ratio of the number of Stage~III models to all models. From top to bottom: JHK color-color plots, IRAC color-color plots, and IRAC+\mips color-color plots. The reddening vector shows an extinction of $\av=20$, assuming the \citet{indebetouw05} extinction law. The red lines in the JHK color-color plots show the locus of the reddened stellar photospheres. The blue lines outline the regions shown in Figure~\ref{f:schematic}.\label{f:op}}
\end{figure}

\clearpage

\begin{figure}
\epsscale{1.0}
\plotone{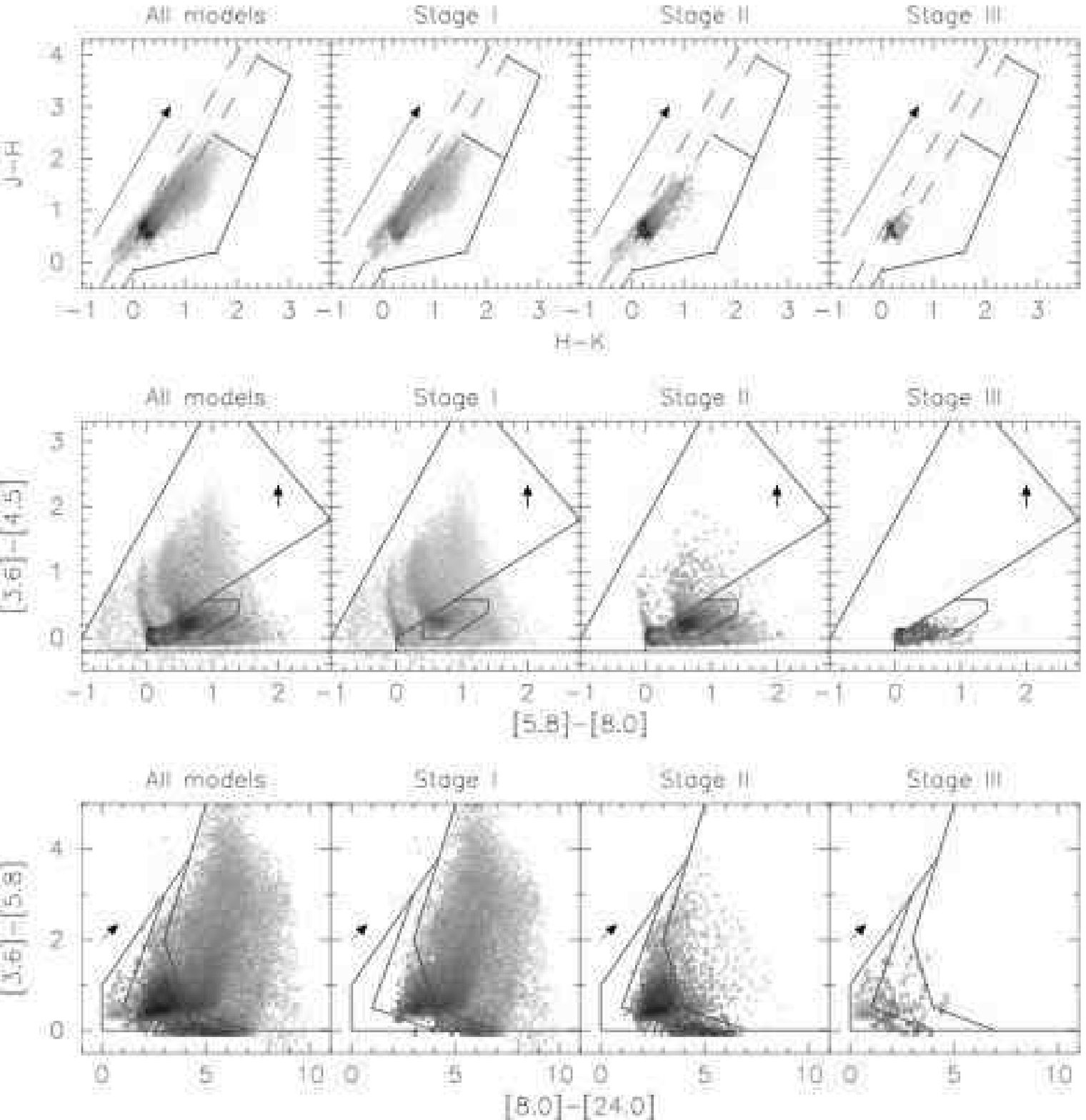}
\caption{As for Figure~\ref{f:o1} for the virtual cluster at $250$\,pc including sensitivity and saturation cutoffs.\label{f:c11}}
\end{figure}

\clearpage

\begin{figure}
\epsscale{0.8}
\plotone{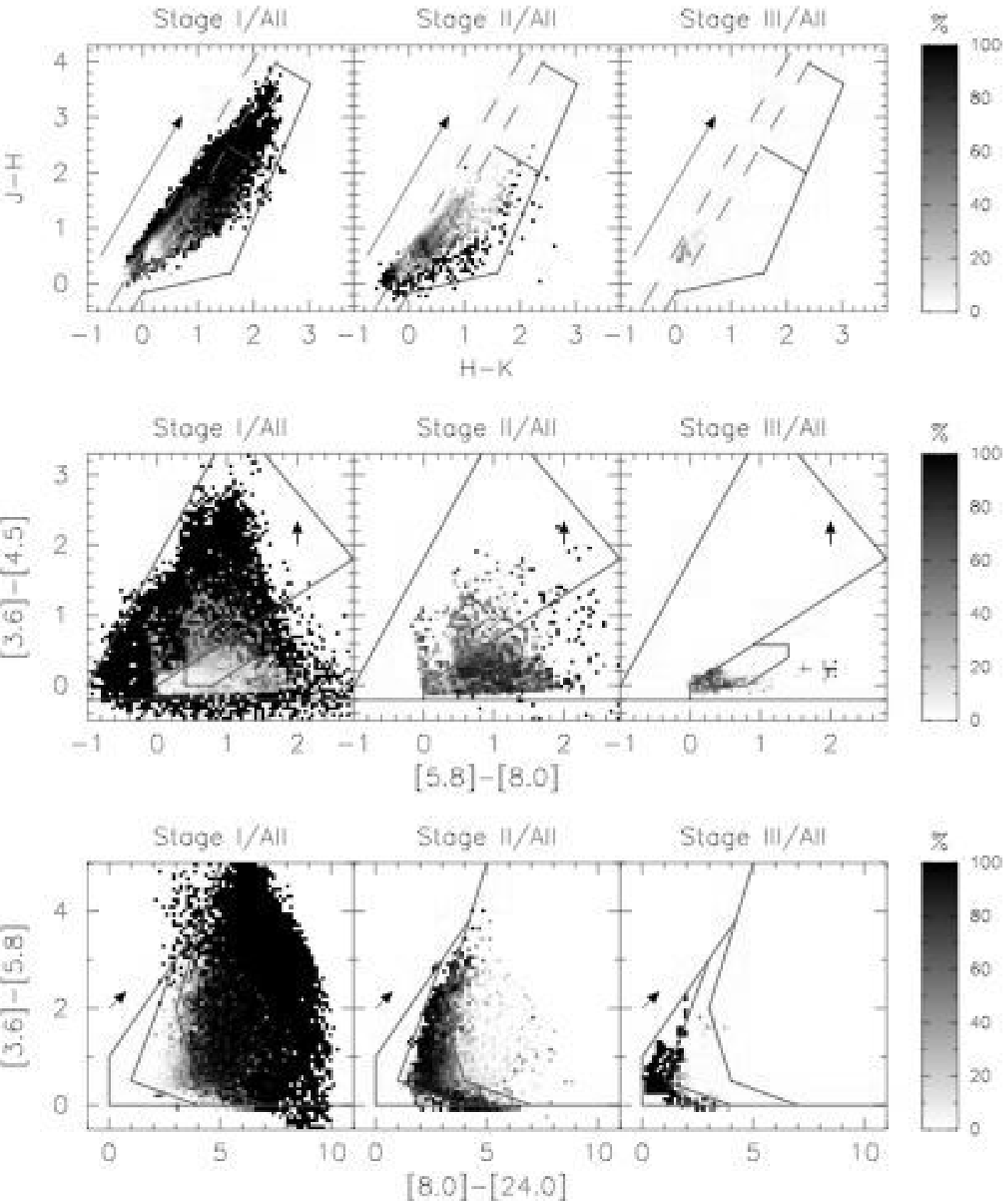}
\caption{As for Figure~\ref{f:op} for the virtual cluster at $250$\,pc including sensitivity and saturation cutoffs.\label{f:c1p}}
\end{figure}

\clearpage

\begin{figure}
\epsscale{1.0}
\plotone{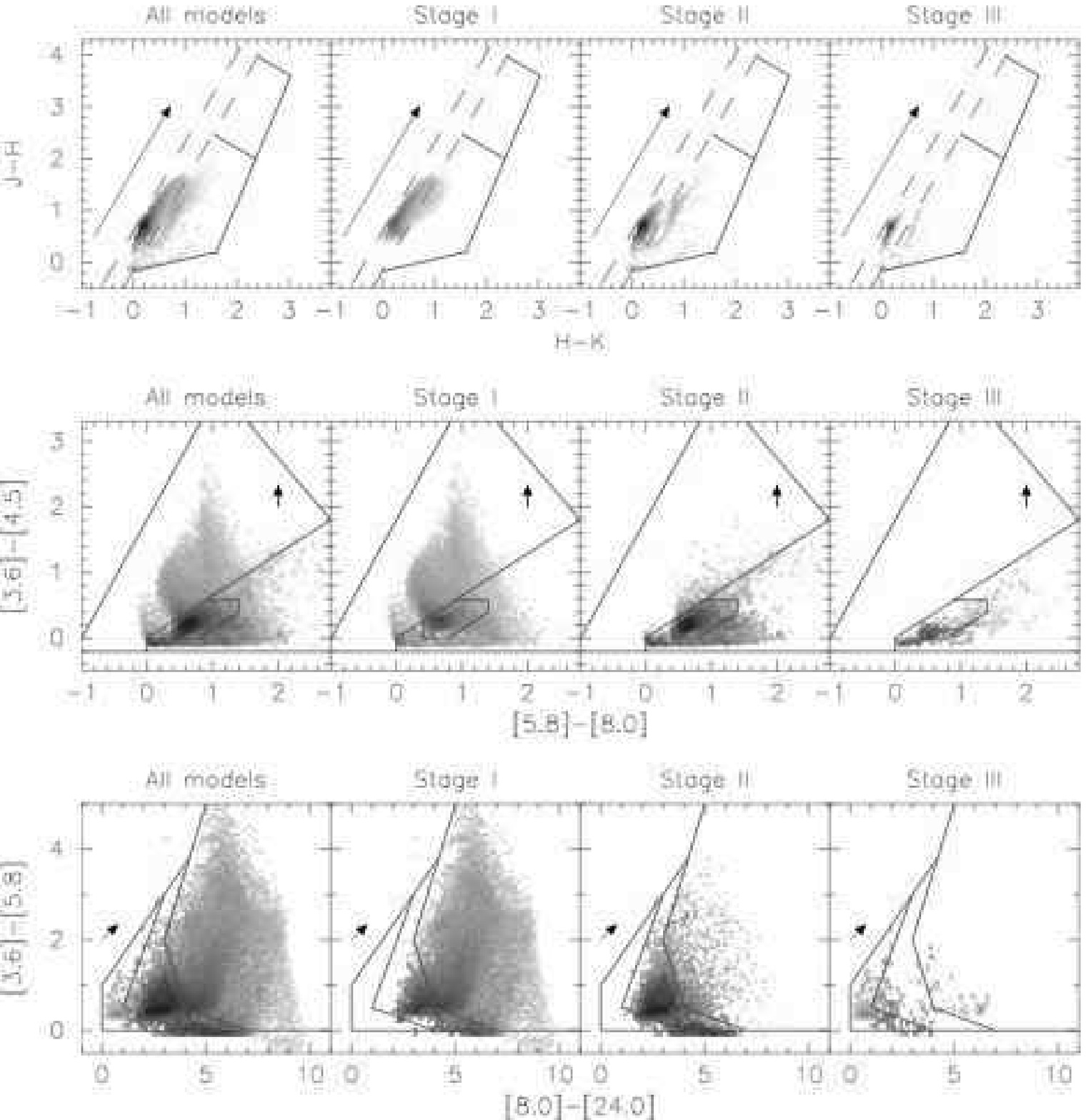}
\caption{As for Figure~\ref{f:o1} for the virtual cluster at $2.5$\,kpc including sensitivity and saturation cutoffs.\label{f:c21}}
\end{figure}

\clearpage

\begin{figure}
\epsscale{0.8}
\plotone{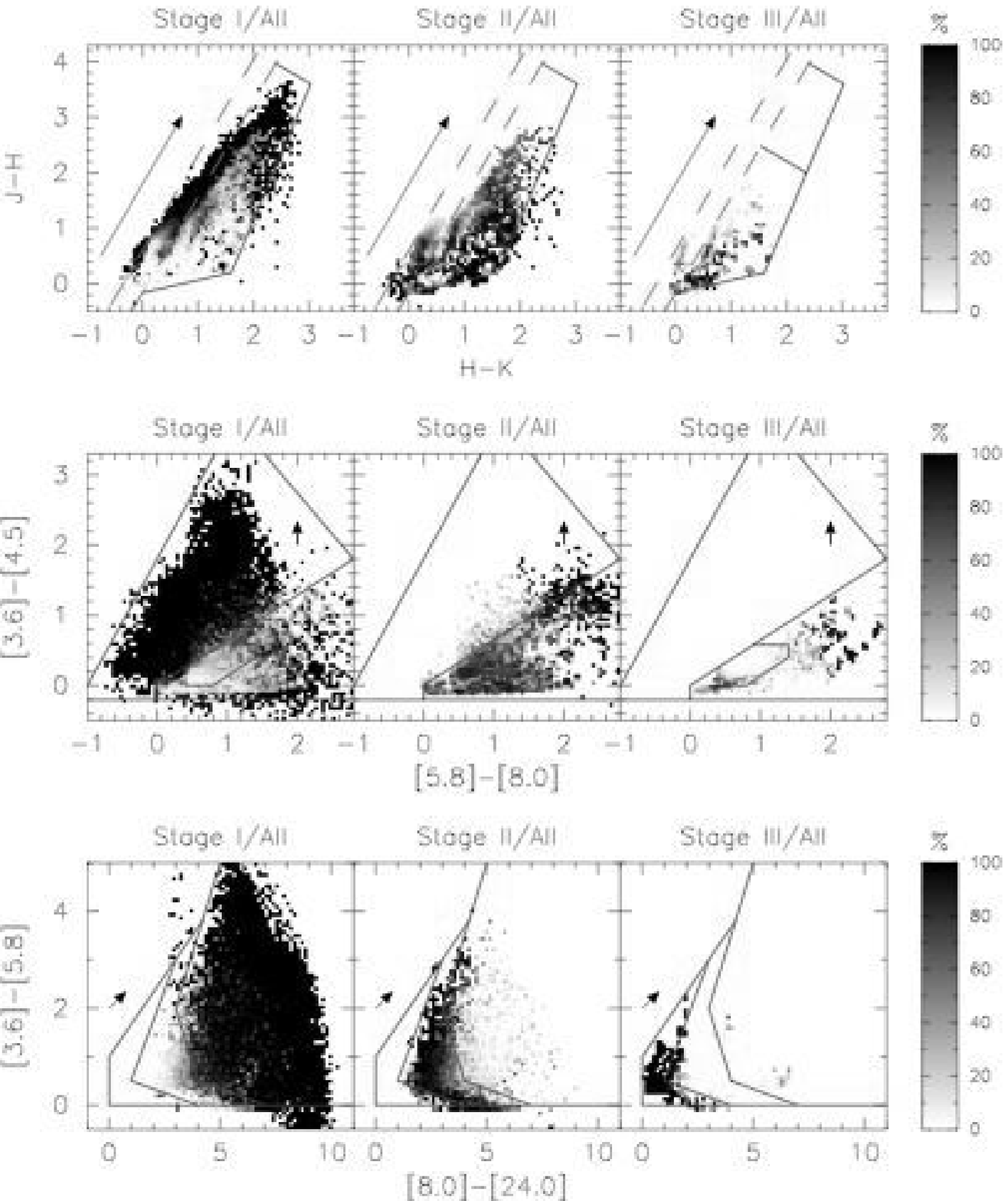}
\caption{As for Figure~\ref{f:op} for the virtual cluster at $2.5$\,kpc including sensitivity and saturation cutoffs.\label{f:c2p}}
\end{figure}

\clearpage

\begin{figure}
\epsscale{1.0}
\plotone{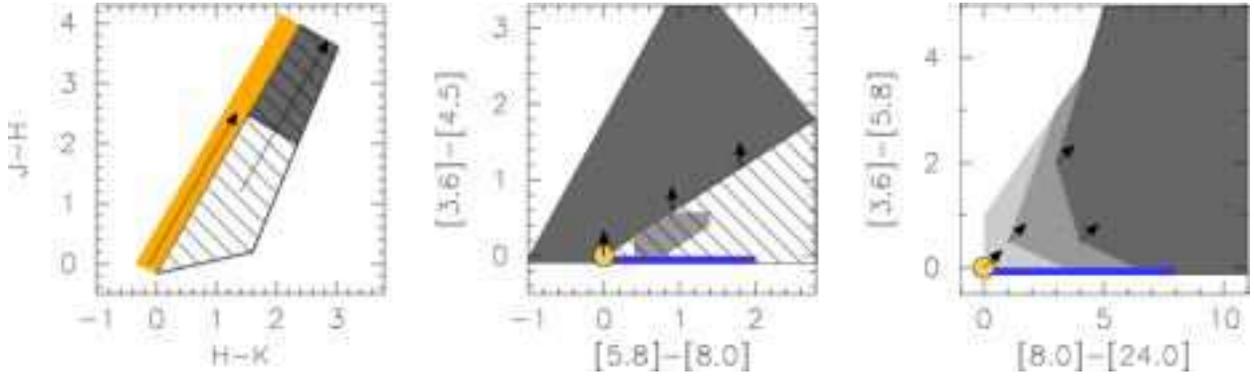}
\caption{Approximate regions of JHK (left), IRAC (center) and IRAC+\mips (right) color-color space where the different evolutionary stages lie. From dark to light gray: the regions where most models are Stage~I, II, and III respectively.
The hashed region in the JHK and IRAC color-color plots are regions where models of all evolutionary stages can be present.
The dark gray region in the JHK color-color plot is also hashed to indicate that this region, although never occupied by Stage~II and III models in the absence of extinction, would easily be contaminated by Stage~II and III models for high values of extinction.
The Stage II area in the IRAC color-color plot is hashed to show that although most models in this region are Stage~II models, Stage~I models can also be found with these colors.
The orange region in the JHK color-color plot represents the location of reddened stellar photospheres.
The yellow disk in the IRAC and IRAC+\mips color-color plots represents the approximate location of stellar photospheres in the absence of extinction.
The blue rectangles show the approximate regions where only disks with large inner holes lie.
The reddening vectors show an extinction of $\av=20$, assuming the \citet{indebetouw05} extinction law.
\label{f:schematic}}
\end{figure}

\clearpage

\begin{figure}
\epsscale{1.0}
\plotone{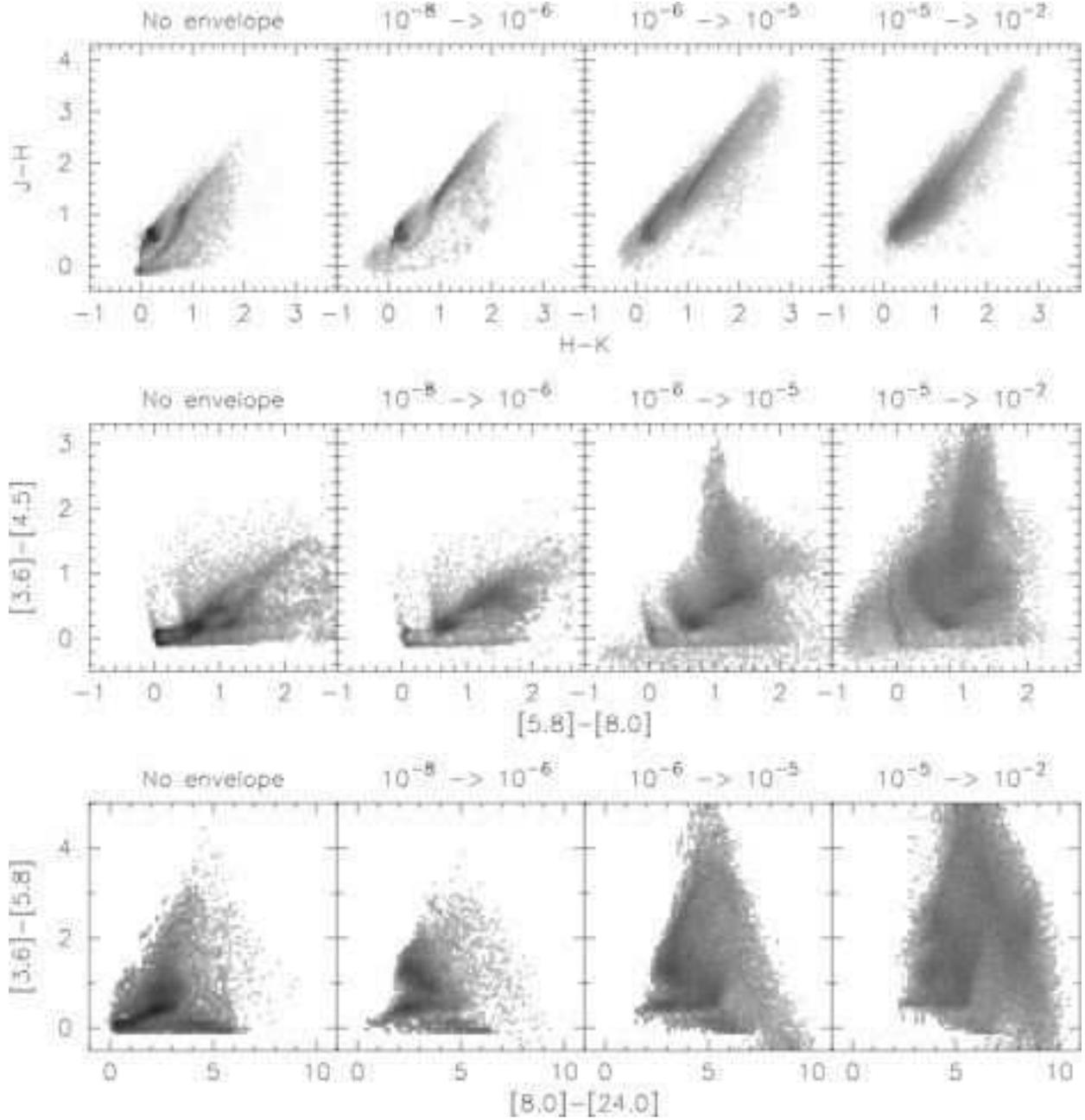}
\caption{The dependence of the JHK, IRAC, and IRAC+\mips colors on the envelope accretion rate $\mdote/\mstar$ (the values are shown above each box; the unit is yr$^{-1}$).
\label{f:cc_mdote}}
\end{figure}

\clearpage

\begin{figure}
\epsscale{1.0}
\plotone{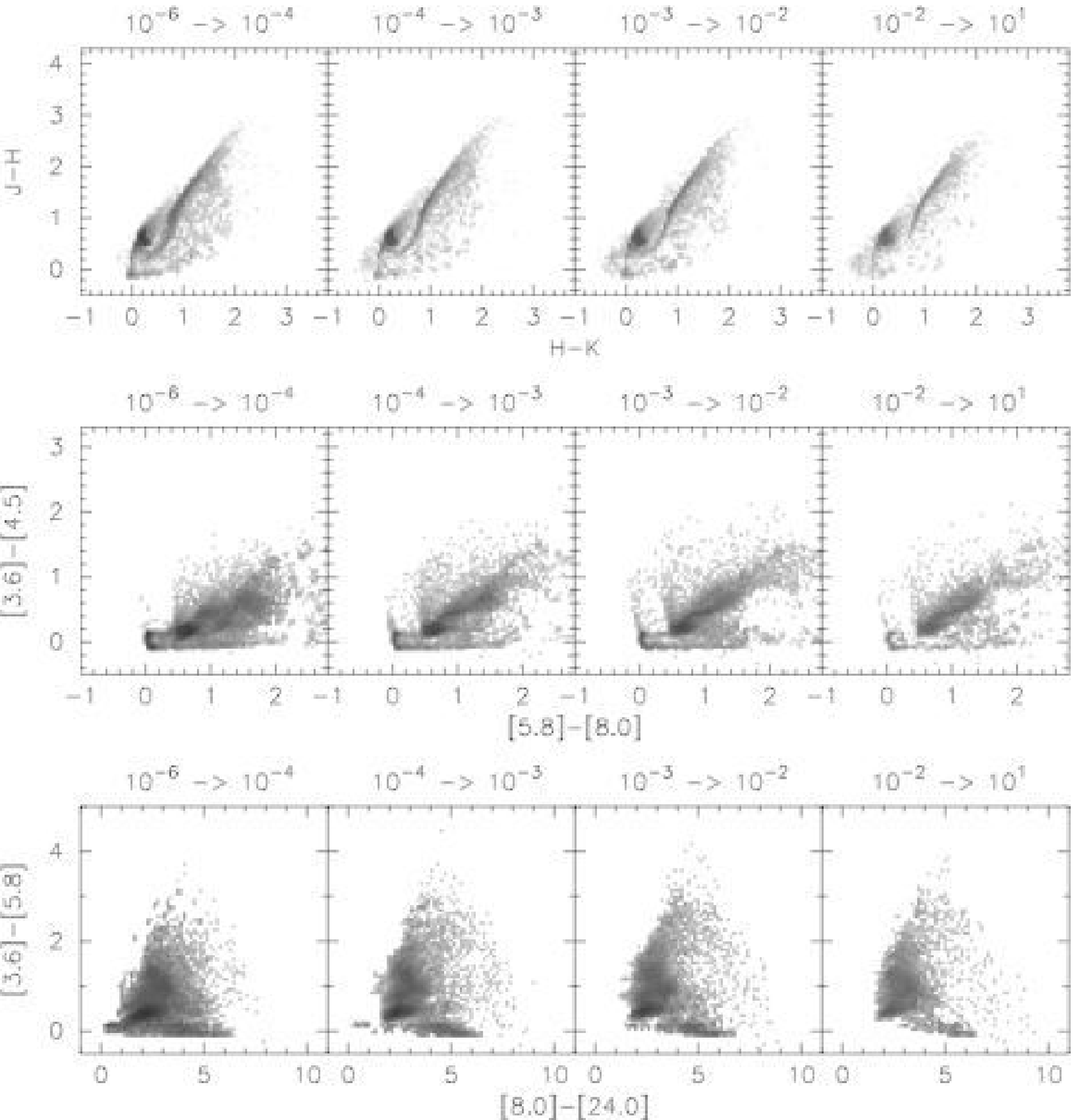}
\caption{The dependence of the JHK, IRAC, and IRAC+\mips colors on the disk mass $\mdisk/\mstar$ (the values are shown above each box). All Stage~II models in the model grid are shown.\label{f:cc_mdisk}}
\end{figure}

\clearpage

\begin{figure}
\epsscale{1.0}
\plotone{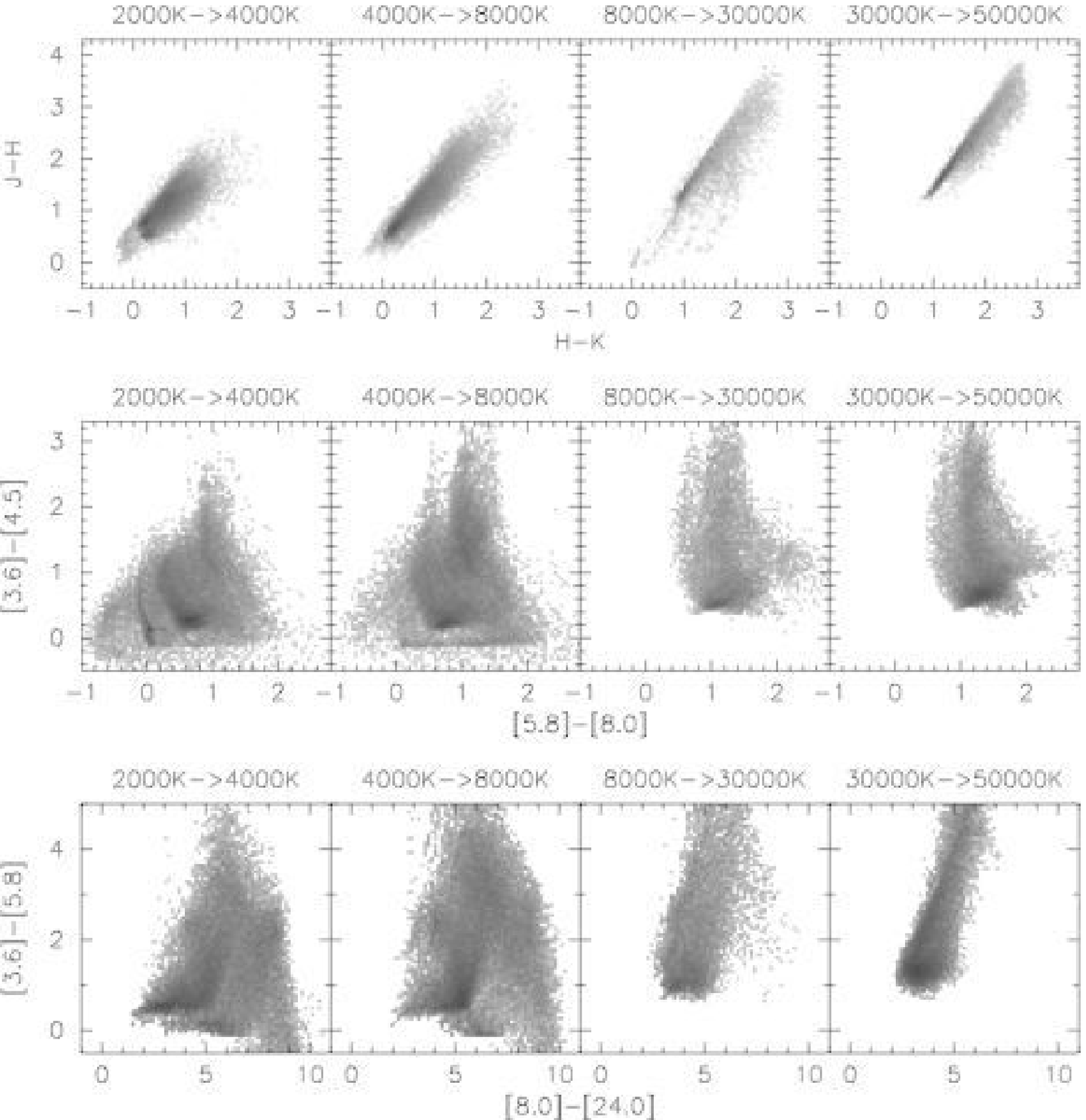}
\caption{The dependence of the JHK, IRAC, and IRAC+\mips colors on the stellar temperature $\tstar$ (the values are shown above each box). All Stage~I models in the model grid are shown.
\label{f:cc_tstar1}}
\end{figure}
\clearpage

\begin{figure}
\epsscale{1.0}
\plotone{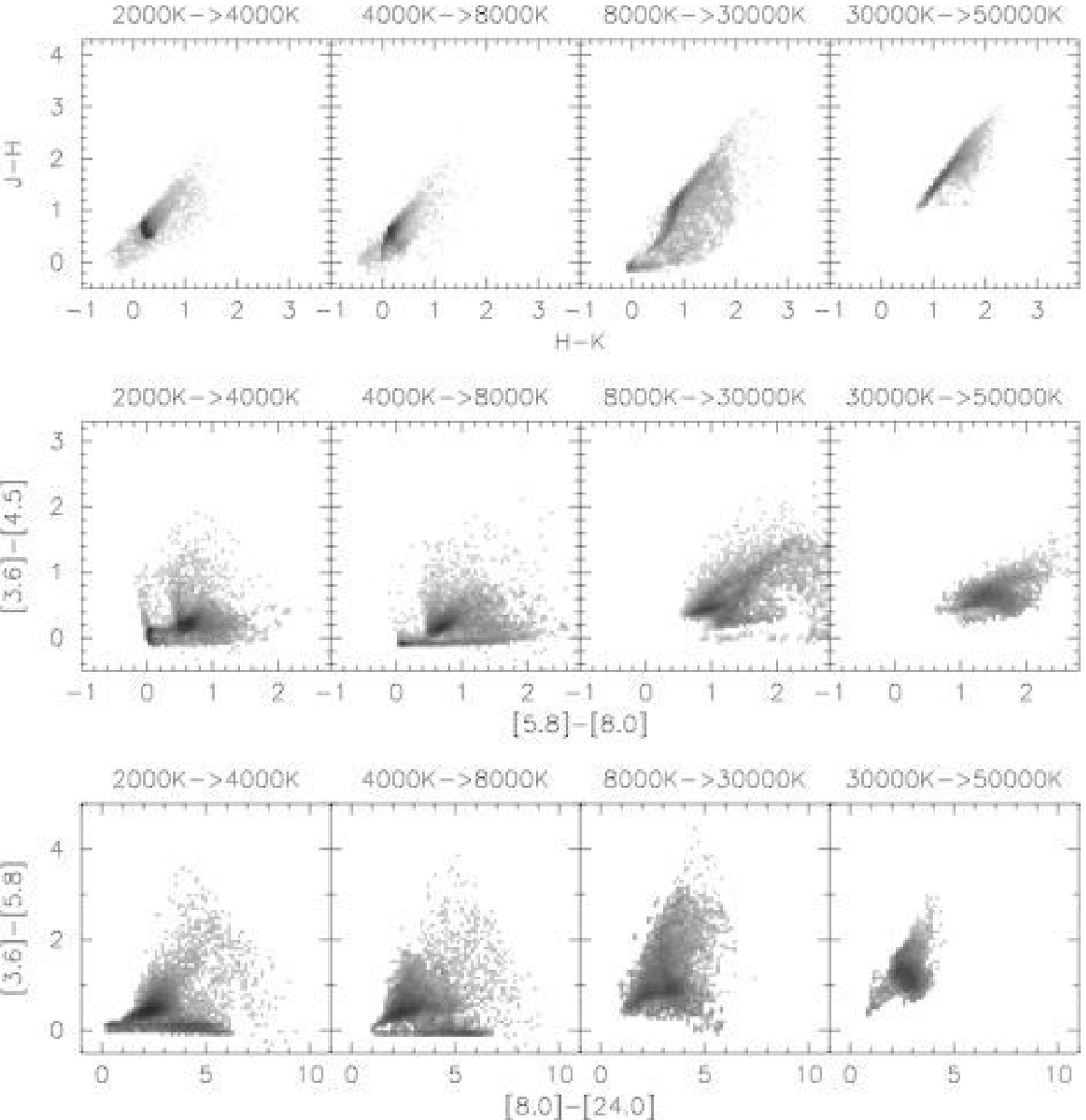}
\caption{The dependence of the JHK, IRAC, and IRAC+\mips colors on the stellar temperature $\tstar$ (the values are shown above each box). All Stage~II models in the model grid are shown.
\label{f:cc_tstar2}}
\end{figure}

\clearpage

\begin{figure}
\epsscale{1.0}
\plotone{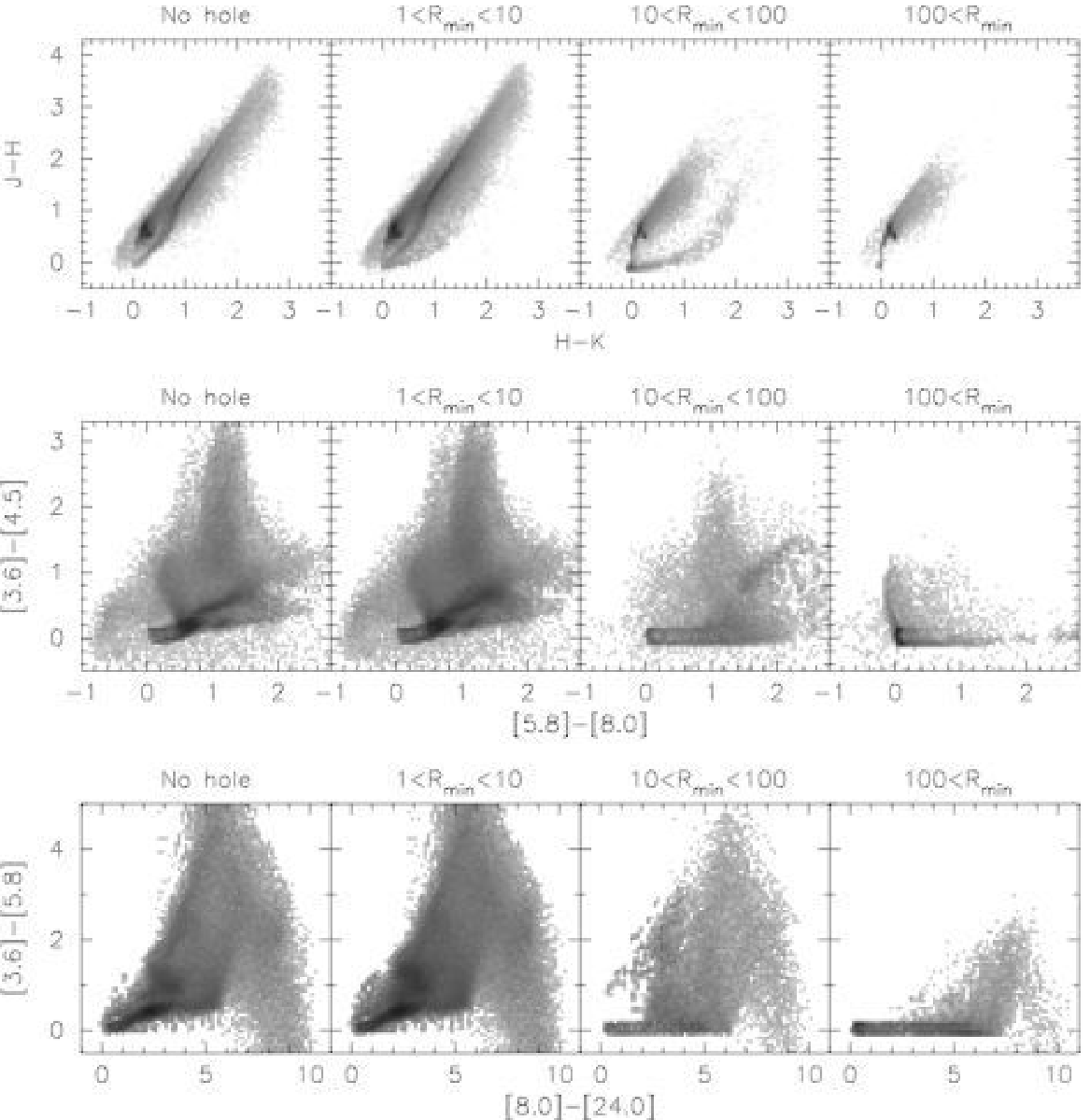}
\caption{The dependence of the JHK, IRAC, and IRAC+\mips colors on the disk and envelope inner radii $\rmind$ ($=\rmine$) (the values are shown above each box - the unit is the dust sublimation radius $\rsub$).
All the models from the grid are shown.
\label{f:cc_rmine}}
\end{figure}

\end{document}